\documentclass[%
 aip,
 jmp,%
 amsmath,amssymb,
 preprint,%
]{revtex4-2}

\usepackage{graphicx}
\usepackage{dcolumn}
\usepackage{bm}

\usepackage{verbatim}
\usepackage{amsmath}

\usepackage{CJKutf8}
\usepackage{epstopdf, epsfig}
\usepackage{xcolor}
\usepackage[caption=false]{subfig}
\usepackage{enumitem}

\newcommand{\rmd}{\mathrm{d}}
\newcommand{\rmi}{\mathrm{i}{\,}}
\newcommand{\rmRe}{\mathrm{Re}}

\begin{document}

\preprint{\emph{Physics of Fluids} \textbf{32}, 126105 (2020); doi.org/10.1063/5.0033933}


\title{Analytical solution for an acoustic boundary layer around an oscillating rigid sphere}

\author{Evert Klaseboer}
\affiliation{%
 Institute of High Performance Computing, 1 Fusionopolis Way, Singapore 138632, Singapore
}%

\author{Qiang Sun(孙强)}
 \email{qiang.sun@rmit.edu.au}
\affiliation{
 Australian Research Council Centre of Excellence for Nanoscale BioPhotonics, School of Science, RMIT University, Melbourne, VIC 3001, Australia
}%

\author{Derek Y. C. Chan}
 \email{D.Chan@unimelb.edu.au}
 \homepage{http:D.Chan.is}
\affiliation{%
School of Mathematics and Statistics, University of
Melbourne, Parkville, VIC 3010, Australia
}%
\affiliation{
 Department of Mathematics, Swinburne University of
Technology, Hawthorn, VIC 3121, Australia
}%


\date{\today}

\begin{abstract}
Analytical solutions in fluid dynamics can be used to elucidate the physics of complex flows and to serve as test cases for numerical models. In this work, we present the analytical solution for the acoustic boundary layer that develops around a rigid sphere executing small amplitude harmonic rectilinear motion in a compressible fluid. The mathematical framework that describes the primary flow is identical to that of wave propagation in linearly elastic solids, the difference being the appearance of complex instead of real valued wave numbers. The solution reverts to well-known classical solutions in special limits: the potential flow solution in the thin boundary layer limit, the oscillatory flat plate solution in the limit of large sphere radius and the Stokes flow solutions in the incompressible limit of infinite sound speed. As a companion analytical result, the steady second order acoustic streaming flow is obtained. This streaming flow is driven by the Reynolds stress tensor that arises from the axisymmetric first order primary flow around such a rigid sphere. These results are obtained with a linearization of the non-linear Navier-Stokes equations valid for small amplitude oscillations of the sphere. The streaming flow obeys a time-averaged Stokes equation with a body force given by the Nyborg model in which the above mentioned primary flow in a compressible Newtonian fluid is used to estimate the time-averaged body force. Numerical results are presented to explore different regimes of the complex transverse and longitudinal wave numbers that characterize the primary flow.

\comment{
\begin{description}
\item[Usage]
Secondary publications and information retrieval purposes.
\item[Structure]
You may use the \texttt{description} environment to structure your abstract;
use the optional argument of the \verb+\item+ command to give the category of each item. 
\end{description}}
\end{abstract}


\begin{CJK*}{UTF8}{gbsn}

\maketitle

\end{CJK*}


\section{Introduction} \label{sec:intro}

Acoustic streaming is the steady flow generated by periodic small amplitude Rayleigh acoustic fields in compressible Newtonian fluids. In certain flow regimes, the streaming patterns are observed as intricate steady vortices or circulatory patterns near solid boundaries. The creation of a steady streaming flow due to a periodic acoustic wave is inherently a non-linear effect that arises from the time averaged velocity field governed by the time-dependent Navier-Stokes equations. The role of fluid viscosity and the generation of vorticity near solid boundaries are key to the streaming phenomenon so that details of such flow depend on the geometry of the system and on the hydrodynamic boundary conditions. Consequently, the flow characteristics will vary near solid or soft surfaces such as biological cell membranes, bubble surfaces or fluid interfaces and as such can provide very different modes of steady fluid transport. Recent advances in microfluidics and ultrasonic technologies stimulated resurgent interest in these phenomena~\citep{bookLaurellLenshof,Bruus2015,Lei2016,Doinikov2018,Karlsen2018,Bach2020,Subbotin2019,Pandey2019}.

From the point of view of analytical and numerical analysis, acoustic streaming has also attracted renewed interest recently, for example around spheres, bubbles and drops \cite{Baasch2020,Otto2008}, including thermal effects~\cite{KarlsenPRE2015,Marshall2015}. Previous studies of steady streaming around a stationary sphere considered an imposed external oscillatory flow field~\citep{Lane1955, Riley1966}, or streaming around bubbles and drops~\cite{Davidson1971,LonguetHiggins1998}. The amplitude of the oscillations is also assumed to be small compared to the sphere radius. As the fluid is generally taken to be incompressible this is termed steady streaming rather than acoustic streaming \citep{Riley2001}. Streaming around a stationary sphere in a compressible fluid has been considered by Lee and Wang~\cite{Lee1990}. However, they made a further assumption to only consider the flow outside the boundary layer to simplify the analysis. 

To elucidate the physics of such complex flows, it is instructive to have available analytical solutions for special cases that can also serve as test cases for more complex numerical models. It turns out that the governing equations and boundary conditions for an oscillating rigid sphere in an infinite but otherwise quiescent compressible Newtonian fluid is very similar to a rigid sphere undergoing oscillatory motion in an infinite linear elastic material for which a analytical solution has been given recently by Klaseboer et al.~\cite{KlaseboerJElas2018}.

This oscillatory motion of a sphere in a compressible Newtonian fluid generates both an acoustic field due to compressibility of the fluid and an acoustic boundary layer in the fluid near the sphere due to viscosity. Without further approximations, we can construct an analytic solution that accounts for the transition from viscous boundary layer dominated flow near interfaces to near potential flow far from boundaries. The known limits of negligible viscosity, negligible compressiblity or zero frequency can be recovered as special cases. In the geometric limit of a large sphere, results for both normal and tangential flows at a planar surface are obtained at different parts of the sphere. To the best knowledge of the authors, such a theoretical solution has not appeared in the literature before. Taking this analytical solution to be the primary solution, we then obtain an expression for the secondary flow or acoustic streaming that originates from small non-linear inertial effects.

At low amplitudes of oscillation~\citep{Stuart1966}, the streaming flow can be obtained as a non-linear steady secondary correction to the linear time-dependent primary flow. In this paper, we present a general analytic solution of the model~\cite{NyborgJASA1953} for streaming flow in a compressible Newtonian fluid around a rigid sphere with radius, $a$ that is executing rectilinear oscillatory motion with angular frequency, $\omega \equiv 2 \pi f$ and velocity amplitude, $U_0$. We focus on the Rayleigh~\citep{bookRayleigh} or acoustic limit in which the magnitude of the sphere displacement, $U_0/\omega$ is small compared to the sphere radius, $a$ and also on the low Reynolds number regime where the non-linear inertial term in the Navier-Stokes equation is small.  

In Section~\ref{sec:nyborg_model} we recapitulate the Nyborg formulation for the steady acoustic streaming flow as a second order effect driven by a linear primary flow. In Section~\ref{sec:formal_soln}, symmetry arguments pertaining to the periodic primary flow due to a rigid sphere executing rectilinear oscillations are used to construct the time-averaged Reynolds stress that results in a steady body force in a Stokes equation that governs the steady streaming velocity field. Possible acoustic waves inside the rigid sphere are not taken into account. An explicit analytic solution for periodic primary flow is given in Section~\ref{sec:Analogy}. Corresponding general solutions for the steady acoustic streaming flow (using the theory of electrophoresis of a charged spherical colloidal particle~\citep{Overbeek1941, JAYARAMAN2019845}) are outlined in Section \ref{sec:SolutionOutline} and are further worked out in Appendix~\ref{app:SecondaryFlowDerivation}. The solutions for the streaming vorticity and velocity are expressed explicitly as integrals of the body force. Analytic and numerical results for the primary flow around a rigid sphere are given in Section~\ref{sec:streaming_examples}. The reduction of this general solution to special cases and geometric limits, together with numerical comparisons are detailed in Section~\ref{sec:classicalsol}, and the concluding remarks are given in Section~\ref{sec:conclusion}. 
%
%
%
\section{The Nyborg framework} \label{sec:nyborg_model}

The Nyborg framework~\citep{NyborgJASA1953}, based on the earlier Eckart theory \citep{Eckart1948}, describes the transmission of an acoustic wave in a compressible Newtonian fluid with constant shear viscosity, $\mu$ and bulk viscosity, $\mu_B$. The governing equations for the space and time dependent density, $\overline \rho(\boldsymbol x,t)$,  velocity field, $\overline {\boldsymbol  v}(\boldsymbol x,t)$ and pressure, $\overline p(\boldsymbol x,t)$ are the continuity and momentum equations
\begin{subequations} \label{eq:cont_mtm_eqn}
\begin{align}
  \frac{\partial \overline \rho} {\partial t} +  \nabla \cdot ( \overline \rho \; \overline {\boldsymbol  v}) = & 0, \\
  \frac{\partial (\overline \rho \; \overline  {\boldsymbol  v}) } {\partial t} + \nabla \cdot (\overline \rho \; \overline {\boldsymbol  v} \; \overline {\boldsymbol  v}) = & - \nabla \overline p + \mu \nabla^2 \overline {\boldsymbol  v}  + \Big(\mu_B + \frac{1}{3} \mu \Big)\nabla (\nabla \cdot \overline {\boldsymbol  v}).
\end{align}
\end{subequations} 
At small vibrating amplitudes, for which all physical quantities can be linearised about their equilibrium values in terms of the small parameters: $\epsilon = |\nabla \cdot (\overline \rho \; \overline {\boldsymbol  v} \; \overline {\boldsymbol  v})| / |{\partial (\overline \rho \; \overline  {\boldsymbol  v}) } / {\partial t}|\sim U_0/(a \omega) \ll 1$, and the Reynolds number, $Re = |\nabla \cdot (\overline \rho \; \overline {\boldsymbol  v} \; \overline {\boldsymbol  v})|/|\mu \nabla^2 \overline {\boldsymbol  v}| \sim \rho_0 a U_0/\mu \ll 1$, all quantities in (\ref{eq:cont_mtm_eqn}) are expanded in powers of $\epsilon$ about the constant reference density $\rho_0$, and pressure, $p_0$ and noting that the reference velocity is zero
\begin{equation}
  \overline \rho = \rho_0 + \epsilon \; \overline{\rho}_1 + \epsilon^2 \; \overline{\rho}_2 + ... , \quad 
  \overline p = p_0 + \epsilon \; \overline{p}_1 + \epsilon^2 \; \overline{p}_2 + ... , \quad
  \overline {\boldsymbol  v} = \epsilon \; \overline{{\boldsymbol  v}}_1 + \epsilon^2 \; \overline{{\boldsymbol  v}}_2 + ... 
\end{equation}
To order $\epsilon$, we have the equations that govern the primary flow
\begin{subequations} \label{eq:order_1}
\begin{align}
  \frac{\partial \overline{\rho}_1} {\partial t} +  \rho_0 \nabla \cdot  \overline {\boldsymbol  v}_1  = & 0,\\
  \rho_0 \frac{\partial \overline {\boldsymbol  v}_1 } {\partial t}   = & - \nabla \overline{p}_1 + \mu \nabla^2 \overline {\boldsymbol  v}_1  + \Big(\mu_B + \frac{1}{3} \mu \Big)\nabla (\nabla \cdot \overline {\boldsymbol  v}_1).
\end{align}
\end{subequations}
For the case of the primary velocity field $\overline {\boldsymbol v}_1(\boldsymbol x, t)$ that is driven by a sphere of radius, $a$ executing rectilinear oscillatory motion with a centre of mass velocity: $\boldsymbol U_0 e^{-\rmi \omega t} = U_0 e^{-\rmi \omega t} \boldsymbol e_z$, along the $z$-direction in a compressible Newtonian fluid, we now show that this primary flow can be obtained analytically because of axial symmetry. We assume harmonic time dependence in all primary flow quantities: $\overline \rho_1(\boldsymbol x, t) \sim \rho(\boldsymbol x) e^{-\rmi \omega t}$, $\overline p_1(\boldsymbol x, t) \sim p(\boldsymbol x) e^{-\rmi \omega t}$ and $\overline {\boldsymbol v}_1(\boldsymbol x, t) \sim \boldsymbol u(\boldsymbol x) e^{-\rmi \omega t}$ so that the order $\epsilon$ equations (\ref{eq:order_1}) become
\begin{subequations} \label{eq:order_1_omega}
\begin{align}
  -\rmi \omega \rho +  \rho_0 \nabla \cdot \boldsymbol u  = & 0,\\[3pt]
   -\rmi \omega \rho_0 \boldsymbol u    = & - \nabla p + \mu \nabla^2 \boldsymbol u  + \Big(\mu_B + \frac{1}{3} \mu \Big)\nabla (\nabla \cdot \boldsymbol  u).
\end{align}
\end{subequations}
For small amplitude acoustic waves, we can assume the equation of state: $\nabla p(\boldsymbol x) = c_0^2 \nabla \rho(\boldsymbol x)$ where $c_0 > 0$ is the constant speed of sound in the fluid. This assumes adiabatic conditions hold~\cite{GopinathTrinh2000}. The pressure, $p$ can thus be eliminated from (\ref{eq:order_1_omega}) to give~\citep{Dual2013Conf}
\begin{equation}  \label{eq:u_eqn}
    [(k_T^2/k_L^2)-1]\nabla (\nabla \cdot \boldsymbol u)+\nabla^2 \boldsymbol u +k_T^2 \boldsymbol u= \boldsymbol 0,
\end{equation} 
with (complex) transverse, $k_T$ and longitudinal, $k_L$ wave numbers defined by
\begin{equation}  \label{eq:kL_and_kT}
k_T^2  \equiv  \rmi \frac{\rho_0 \omega}{\mu}  \quad \text{and} \quad k_L^2  \equiv   \displaystyle{\omega^2 \Big/ \left[ c_0^2 - \frac{\rmi \omega}{\rho_0 }\Big(\mu_B +\frac{4}{3}\mu\Big) \right]}.
\end{equation}
Note that $|k_L^2| \leq (3/4)|k_T^2|$ and that the viscous penetration depth defined as $\delta=\sqrt{2 \mu/(\rho_0 \omega)}$ \citep{bookSchlichting} is closely related to the Womersley number $M$~\cite{Riley1966, SadhalBook2015} as $|M|=\sqrt{2} a / \delta =|k_T|a$.  In Sec.~\ref{sec:formal_soln}, Eq.~(\ref{eq:u_eqn}) will be solved for a rigid sphere executing periodic rectilinear motion with no-slip fluid boundary conditions, such that boundary layers are specifically taken into account.


\section{Formal solution for axisymmetric flow} \label{sec:formal_soln}

In this section we exploit the axial symmetry condition of the primary flow driven by the rectilinear motion of a rigid sphere to construct a general formal solution of the problem. We draw on a previously obtained solution for a rigid sphere oscillating in an infinite linearly elastic (solid mechanics) domain. 

\subsection{Symmetry of the first order primary flow} \label{sub_sec:formal_1st_order}
The solution of the order $\epsilon$ equation (\ref{eq:u_eqn}) due to the oscillatory motion of a rigid sphere along the  $z$-direction with velocity amplitude, $U_0$ can only depend on the vector $\boldsymbol U_0$ and the position vector $\boldsymbol x$ with the origin at the centre of the sphere. 
Symmetry consideration implies that the solution to (\ref{eq:u_eqn}) has the general form
\begin{equation} \label{eq:u_general}
\begin{aligned}
    \boldsymbol{u}(\boldsymbol x) =& u_r(r) \cos\theta \; \boldsymbol{e}_r+ u_\theta(r) \sin\theta \; \boldsymbol{e}_\theta\\
    =& \left\{-\frac{2}{r}h(r)+\frac{\rmd \phi(r)}{\rmd r} \right\}  U_0\cos \theta \; \boldsymbol e_r+\left\{\frac{1}{r}\frac{\rmd}{\rmd r} \Big[r h(r) \Big]-\frac{\phi(r)}{r} \right\}  U_0 \sin \theta \; \boldsymbol e_
    \theta
\end{aligned}
\end{equation}
where $u_r(r)$ and $u_\theta(r)$ are only functions of the radial distance, $r$ from the centre of the sphere and $\boldsymbol{e}_r$ and $\boldsymbol{e}_\theta$ are unit vectors in the direction of increasing radial and polar coordinates relative to the $z$-direction. In general, a vector field, $\boldsymbol u$ can be expressed as the sum of a divergence free component, $\boldsymbol u_T$ with $\nabla \cdot \boldsymbol u_T = 0$ and an irrotational component, $\boldsymbol u_L$ with $\nabla \times \boldsymbol u_L =  \boldsymbol 0$, as shown in Landau and Lifshitz~\citep[p. 101--106, {\S}22]{bookLandau}. Since $\boldsymbol{u}(\boldsymbol x)$ is independent of the azimuthal angle $\varphi$ and has no components in the $\boldsymbol e_\varphi$ direction, the irrotational longitudinal component of $\boldsymbol u$ can be represented as $\boldsymbol{u}_L(\boldsymbol x) = \nabla [\phi(r)\cos\theta]\, U_0$, where $\phi(r)$ satisfies $(\nabla^2 +k_L^2) [\phi(r)\cos\theta] = 0$. Similarly, the divergence free transverse component of the velocity can be represented as $\boldsymbol{u}_T(\boldsymbol x) = -\nabla \times [h(r) \sin\theta \;  \boldsymbol{e}_\varphi]\, U_0$ with $(\nabla^2 +k_T^2) [h(r)\cos\theta] = 0$. The representation (\ref{eq:u_general}) then follows.

The introduction of $\phi(r)$ and $h(r)$ simplifies the expression for the pressure and the vorticity of the primary flow. From the continuity equation and the equation of state of the compressible fluid, the pressure is:
\begin{equation}\label{eq:primary_p}
    p(\boldsymbol{x}) = \frac{\rho_0 c_{0}^2}{\rmi \omega} \; \nabla \cdot \boldsymbol{u}(\boldsymbol{x}) 
    = \frac{\rho_0 c_0^2}{\rmi \omega} \nabla^2 [\phi(r) \cos \theta]\; U_0  = \frac{\rmi\rho_0 c_0^2}{ \omega} k_L^2 \phi(r) \cos\theta \; U_0.
\end{equation}
The pressure in the incompressible limit where $\nabla \cdot \boldsymbol{u} \rightarrow 0$ and the speed of sound, $c_0 \rightarrow \infty$, reduces to the familiar acoustic result: $p(\boldsymbol{x}) = \rmi \omega \rho_0 \phi(r) \cos\theta \, U_0$. From (\ref{eq:u_general}), we find for the vorticity:
\begin{equation}\label{eq:w}
\boldsymbol{w}(\boldsymbol{x}) \equiv \nabla \times \boldsymbol{u}(\boldsymbol{x})  = - \frac{\rmd }{\rmd {\theta}} \left\{  \nabla^2 [h(r) \cos \theta] \right\}  U_0  \; \boldsymbol{e}_{\varphi} = - k_{T}^2 h(r) \sin \theta  \; U_0  \; \boldsymbol{e}_{\varphi}.
\end{equation}

The order $\epsilon$ equation (\ref{eq:u_eqn}) that governs the primary flow, $\boldsymbol u(\boldsymbol{x})$ has the same mathematical form as the equation for elastic waves in linear elasticity if the longitudinal and transverse velocity of the elastic waves are identified formally as: $c_T^2 \equiv \omega^2/k_T^2$ and $c_L^2 \equiv \omega^2/k_L^2$, respectively. This analogy is explored further in Sec.~\ref{sec:Analogy}.

\subsection{The analogy with dynamic linear elasticity}\label{sec:Analogy}
In a linear dynamic elastic system, equilibrium of forces requires $\nabla \cdot \boldsymbol \Sigma = \rho \partial^2 \boldsymbol U / \partial t^2$, with $\boldsymbol \Sigma$ the stress tensor, $\boldsymbol U$ the displacement, $\rho$ the density and $t$ time. Assuming harmonic motion with angular frequency $\omega$, one can write $\boldsymbol \Sigma = \boldsymbol \sigma e^{-\rmi\omega t}$ and $\boldsymbol U = \boldsymbol u e^{-\rmi\omega t}$, then (in the frequency domain) the equation of motion becomes $
    \nabla \cdot \boldsymbol \sigma = - \rho \omega^2 \boldsymbol u$.
Linear isotropic homogeneous materials satisfy Hooke's law as:
\begin{equation}
    \frac{\boldsymbol \sigma}{\mu}=\left[\frac{k_T^2}{k_L^2}-2 \right](\nabla \cdot \boldsymbol u) \boldsymbol I + \nabla \boldsymbol u + [\nabla \boldsymbol u]^T
\end{equation}
with $\mu$ the shear modulus, $\boldsymbol I$ the identity matrix and superscript $^{`T\text{'}}$ indicating the transpose. The longitudinal ($k_L$) and transverse ($k_T$) wave numbers in linear elasticity are defined as $k_L^2=\omega^2 \rho/(\lambda + 2 \mu)$ and $k_T^2=\omega^2 \rho/\mu$, ($\lambda$ is one of the Lam\'{e} constants) and are real valued quantities. When Hooke's law is substituted into the equation of motion, one gets:
\begin{equation}
[(k_T^2/k_L^2)-1] \nabla (\nabla \cdot \boldsymbol u) + \nabla^2 \boldsymbol u + k_T^2 \boldsymbol u = \boldsymbol 0.     
\end{equation}
This equation is identical to (\ref{eq:u_eqn}). Since the boundary conditions are also identical being $\boldsymbol u = \boldsymbol U_0$, then these two systems should have identical mathematical solutions. In Klaseboer et al.~\cite{KlaseboerJElas2018}, an analytical solution was given for the elastic waves emitted by a rigid sphere oscillating in an infinite elastic medium. The solution of the elastic wave problem can be represented as a combination of the Green's function and a dipole field and explicit solutions for the velocity components were given in Klaseboer et al~\cite{KlaseboerJElas2018} as:
\begin{equation}
    \begin{aligned} \label{eq:uLE_xyz}
    \boldsymbol u =c_1 2 \frac{a}{r}\bigg\{ e^{\rmi k_Tr}\bigg( [1+G(k_Tr)]\boldsymbol U_0
    +F(k_Tr) \frac{\boldsymbol x \cdot \boldsymbol U_0}{r^2}\boldsymbol x
    \bigg) \\
    -e^{\rmi k_Lr}\frac{k_L^2}{k_T^2}\bigg(G(k_Lr) \boldsymbol U_0 + F(k_Lr) \frac{\boldsymbol x \cdot \boldsymbol U_0}{r^2} \boldsymbol x \bigg)\bigg\}\\
    -c_2 \frac{a^3}{r^3} e^{\rmi k_Lr}\big\{(\rmi k_Lr-1) \boldsymbol U_0 + k_L^2  F(k_Lr)(\boldsymbol x \cdot \boldsymbol U_0) \boldsymbol x  \big\}
    \end{aligned}
\end{equation}
with the functions $F(y)\equiv -1-3\rmi /y+3/y^2$, $G(y)\equiv \rmi/y -1/y^2$ and $\boldsymbol x$ the position vector with respect to the center of the sphere. This solution will also be valid in the acoustic boundary layer context. Note however that in linear elasticity $k_L$ and $k_T$ are real parameters, while in the current case of acoustic streaming they are complex parameters as given in Eq.~(\ref{eq:kL_and_kT}).

\subsection{Solution for the primary flow} \label{sec:primary_flow}

When an axial symmetric coordinate system is used, using $\boldsymbol U_0 = U_0 \cos \theta \boldsymbol e_r - U_0 \sin \theta \boldsymbol e_\theta$, $\boldsymbol x = r \boldsymbol e_r$ and $(\boldsymbol x \cdot \boldsymbol U_0)=r U_0 \cos \theta$, Eq.~(\ref{eq:uLE_xyz}) becomes:
\begin{subequations} \label{eq:ur_utheta}
\begin{align}
    \frac{u_\theta(r)}{U_0} =& C_1 \frac{a}{r} \; \big[1+G(k_Tr)\big] \; e^{\rmi k_Tr} +  C_2 \; \frac{a}{r} \; G(k_Lr) \; e^{\rmi k_Lr}, \\
    \frac{u_r(r)}{U_0} =&  2 \; C_1 \; \frac{a}{r} \; G(k_Tr) \; e^{\rmi k_Tr}  + C_2 \frac{a}{r} \; [1+2G(k_Lr)] \; e^{\rmi k_Lr}.
\end{align}
\end{subequations}
It is more convenient, instead of the constants $c_1$ and $c_2$ of Eq.~(\ref{eq:uLE_xyz}), to introduce new constants $C_1=-2c_1$ and $C_2=k_L^2 a^2[c_1 2/(k_T^2 a^2)+ c_2]$ in (\ref{eq:ur_utheta}). The functions $h(r)$ and $\phi(r)$ defined in (\ref{eq:u_general}) can now seen to be:
\begin{equation} \label{eq:h_phi}
 h(r)= -C_1  \; a \; G(k_T r) \; e^{\rmi k_Tr}, \quad \quad
 \phi(r)=- C_2 \; a\; G(k_L r) \; e^{\rmi k_L r}.
\end{equation}
From the no-slip boundary conditions at the surface $r=a$: $u_r(a)=U_0$ and $u_\theta(a)=-U_0$ we find:
\begin{equation} \label{eq:C1_C2}
    C_1=-\frac{[1+3G(k_L a)]}{[1+G(k_T a)+2G(k_L a)]} e^{-\rmi k_T a}, \quad  C_2 =  \frac{[1+3G(k_T a)]}{[1+G(k_T a)+2G(k_L a)]} e^{-\rmi k_L a}.
\end{equation}
The complex valued $k_T$ and $k_L$ are taken to have positive imaginary parts to ensure a finite solution at infinity. We can now also identify terms in $\exp(\rmi k_L r)$ and $\exp(\rmi k_T r)$ to be the longitudinal, $\boldsymbol u_L$ and transverse, $\boldsymbol u_T$ components of $\boldsymbol u$ respectively.

The solutions (\ref{eq:ur_utheta}) and (\ref{eq:h_phi}) reduce to potential flow solutions when $|k_T a| \gg 1$, as shown in Section~\ref{sec:potential}. When the radius, $a$ is large, at $\theta=\pi/2$ the Stokes solution for an oscillating plate is recovered and at $\theta=0$ a plane sound wave is recovered, as illustrated in Section~\ref{sec:flat_plate}. In the incompressible limit we recover the result given by Landau and Lifshitz~\citep[p. 89, {\S}24 Problem 5]{bookLandau_FM} and Stokes flow if the frequency $\omega$ tends to zero, as demonstrated in Section~\ref{sec:steady_stream}. The recovery of these classical solutions gives us further confidence that the constructed solution is indeed correct.

If the sphere has a zero tangential stress boundary condition, the vanishing of the tangential stress due to the primary flow will determine different constants $C_1$ and $C_2$. 

\subsection{Solution procedure for the secondary flow} \label{sec:SolutionOutline}

The secondary `streaming' flow will now be obtained, assuming the primary flow is as given in Sec.~\ref{sec:primary_flow}. The streaming velocity is determined by the order $\epsilon^2$ terms of the momentum equation (Eq.~\ref{eq:cont_mtm_eqn}b):
\begin{equation} \label{eq:order_2}
  \rho_0 \frac{\partial \overline {\boldsymbol  v}_2 } {\partial t} +  \frac{\partial (\overline \rho_1 \; \overline  {\boldsymbol  v}_1) } {\partial t} +  \rho_0 \;\nabla \cdot (\overline {\boldsymbol  v}_1 \; \overline {\boldsymbol  v}_1)  =  - \nabla \overline{p}_2 + \mu \nabla^2 \overline {\boldsymbol  v}_2  + \Big(\mu_B + \frac{1}{3} \mu \Big)\nabla (\nabla \cdot \overline {\boldsymbol  v}_2).
\end{equation}
For harmonic time dependence of all quantities: $\overline \phi(\boldsymbol x,t) \sim e^{-\rmi \omega t}$, we define the corresponding steady, time-independent streaming quantities by taking the time average: $ \phi(\boldsymbol x) \equiv <\overline \phi(\boldsymbol x,t)> = (1/T) \int_0^T \overline \phi(\boldsymbol x,t) \, \rmd t$ over one period of oscillation: $T = 1/f = 2\pi/\omega$.

By time-averaging the order $\epsilon^2$ momentum equation  (\ref{eq:order_2}), we obtain the desired Stokes equation relating the time-averaged body force, $\boldsymbol {\mathcal{F}} (\boldsymbol x)$ to the time-average pressure, $P(\boldsymbol x) \equiv <\overline{p}_2(\boldsymbol x,t)>$, and the streaming velocity, $\boldsymbol U(\boldsymbol x) \equiv <\overline {\boldsymbol  v}_2(\boldsymbol x,t)> $:
\begin{equation} \label{eq:streaming_eqn}
    \boldsymbol {\mathcal{F}} (\boldsymbol x) \equiv \rho_0 \;\nabla \cdot <\overline {\boldsymbol  v}_1 \; \overline {\boldsymbol  v}_1>  =  - \nabla P(\boldsymbol x) + \mu \nabla^2 \boldsymbol U(\boldsymbol x).
\end{equation}
Since $\partial (\overline \rho_1 \; \overline  {\boldsymbol  v}_1) /{\partial t}$ and $\partial ( \overline  {\boldsymbol  v}_2) /{\partial t}$ are periodic, their time average is zero. The time averaging also removes the periodic part of $(\overline {\boldsymbol  v}_1 \; \overline {\boldsymbol  v}_1)$, $\overline{p}_2(\boldsymbol x,t)$ and $\overline {\boldsymbol  v}_2(\boldsymbol x,t)$. Furthermore, the streaming velocity, $\boldsymbol U(\boldsymbol x)$ is divergence free, $\nabla \cdot \boldsymbol U = 0$ \citep{NyborgJASA1953, Lee1990}, based on a physical argument that there are no steady sources or sinks. However, a formal proof of this result is not that straightforward. The secondary velocity satisfies $\nabla \cdot \boldsymbol U(\boldsymbol x)=-\nabla \cdot <\overline {\boldsymbol v}_1 \overline \rho_1>/\rho_0$. Thus it appears that both sides of this equation are of order $\epsilon^2$.  While Nyborg~\cite{NyborgJASA1953} assumed a solenoidal secondary flow from the onset, Lee and Wang~\cite{Lee1990} proved more rigorously that $\nabla \cdot \boldsymbol U(\boldsymbol x)=0$ using a low viscosity argument. 

Accepting that the streaming velocity, $\boldsymbol U$ is divergence free, its governing equation (\ref{eq:streaming_eqn}) is identical to a Stokes flow in the presence of a non-conservative body force, $\boldsymbol {\mathcal{F}}$ due to the time-averaged Reynolds stress tensor, $<-\rho_0 (\overline {\boldsymbol  v}_1 \; \overline {\boldsymbol  v}_1)>$ that is given in terms of the first order velocity field, $\overline {\boldsymbol v}_1$. 

This is the starting point for the general small amplitude theory of the streaming velocity by calculating $\boldsymbol {\mathcal{F}}$ from the solution of the primary flow. The desired physical solution is $\overline {\boldsymbol v}_1(\boldsymbol x, t) = \text{Real} \{ \boldsymbol u(\boldsymbol x) e^{-\rmi \omega t} \}$ where $\boldsymbol u(\boldsymbol x) = \boldsymbol u'(\boldsymbol x) + \rmi \boldsymbol u''(\boldsymbol x)$ is in general complex with real and imaginary parts $\boldsymbol u'(\boldsymbol x)$ and $\boldsymbol u''(\boldsymbol x)$ and the time-averaged body force in (\ref{eq:streaming_eqn}) becomes, $\boldsymbol{\mathcal{F}} = {\textstyle\frac{1}{2}} \rho_0 \nabla \cdot [\boldsymbol u' \boldsymbol u' + \boldsymbol u'' \boldsymbol u'']$. 

Drawing on the mathematical concepts used in electrophoresis of a spherical particle, explicit solutions for the streaming vorticity and velocity fields can be given in terms of the body force. The derivation is rather tedious but otherwise relatively straightforward and is provided in Appendix~\ref{app:SecondaryFlowDerivation}.

\section{Examples of the primary velocity field} \label{sec:streaming_examples}

\begin{figure*} 
\centering{}
\subfloat[]{ \includegraphics[width=0.42\textwidth]{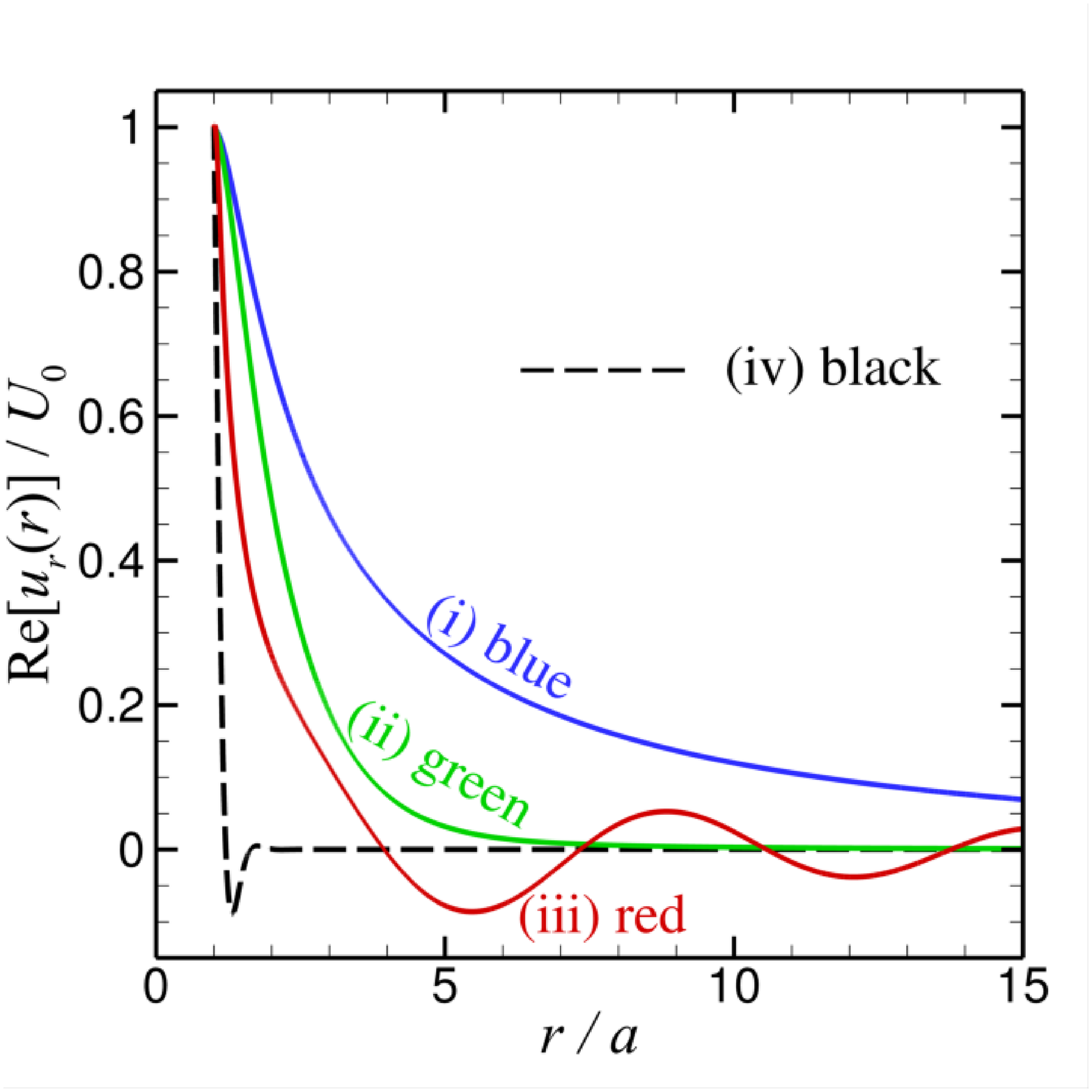}}
\subfloat[]{ \includegraphics[width=0.42\textwidth]{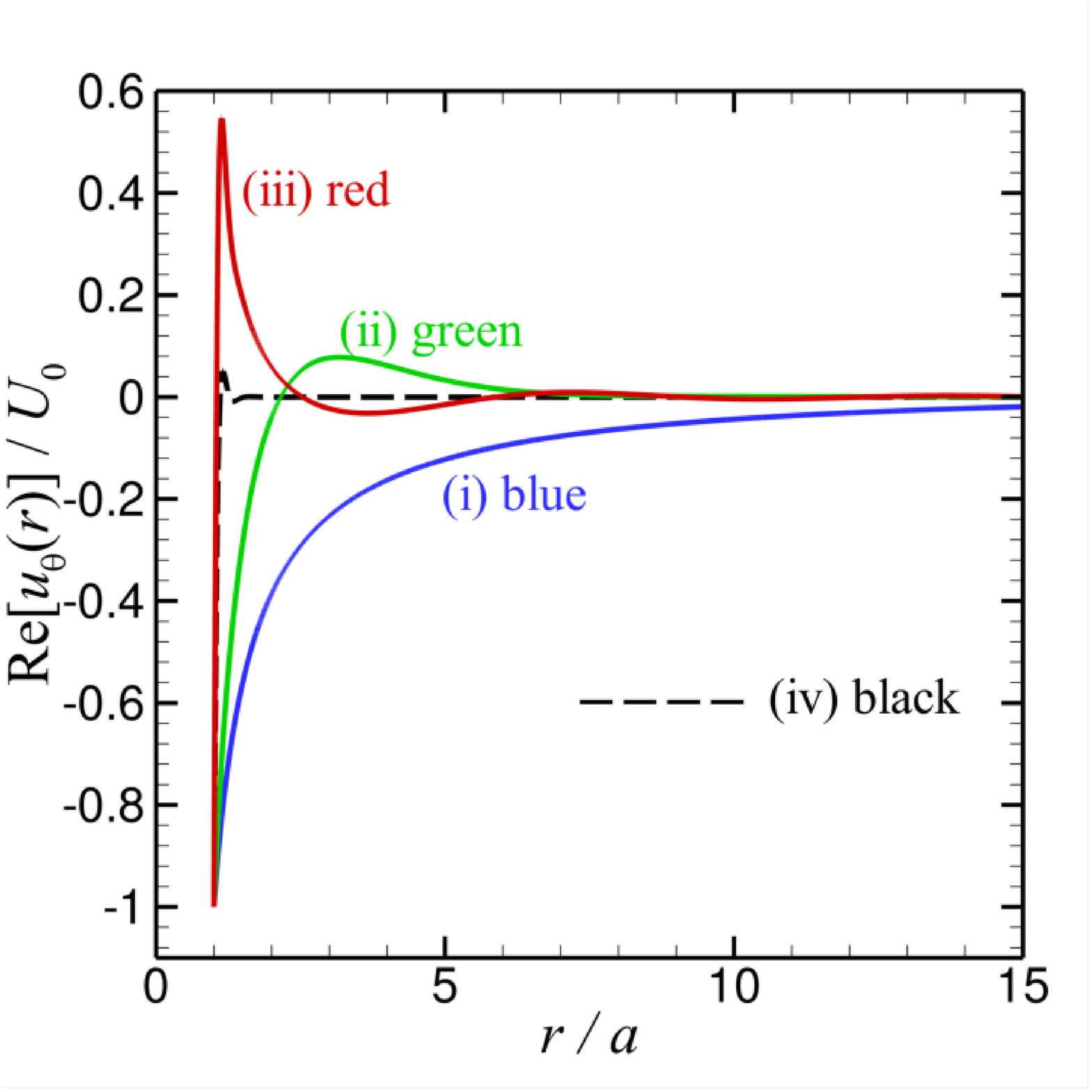}}
\\
\subfloat[]{ \includegraphics[width=0.42\textwidth]{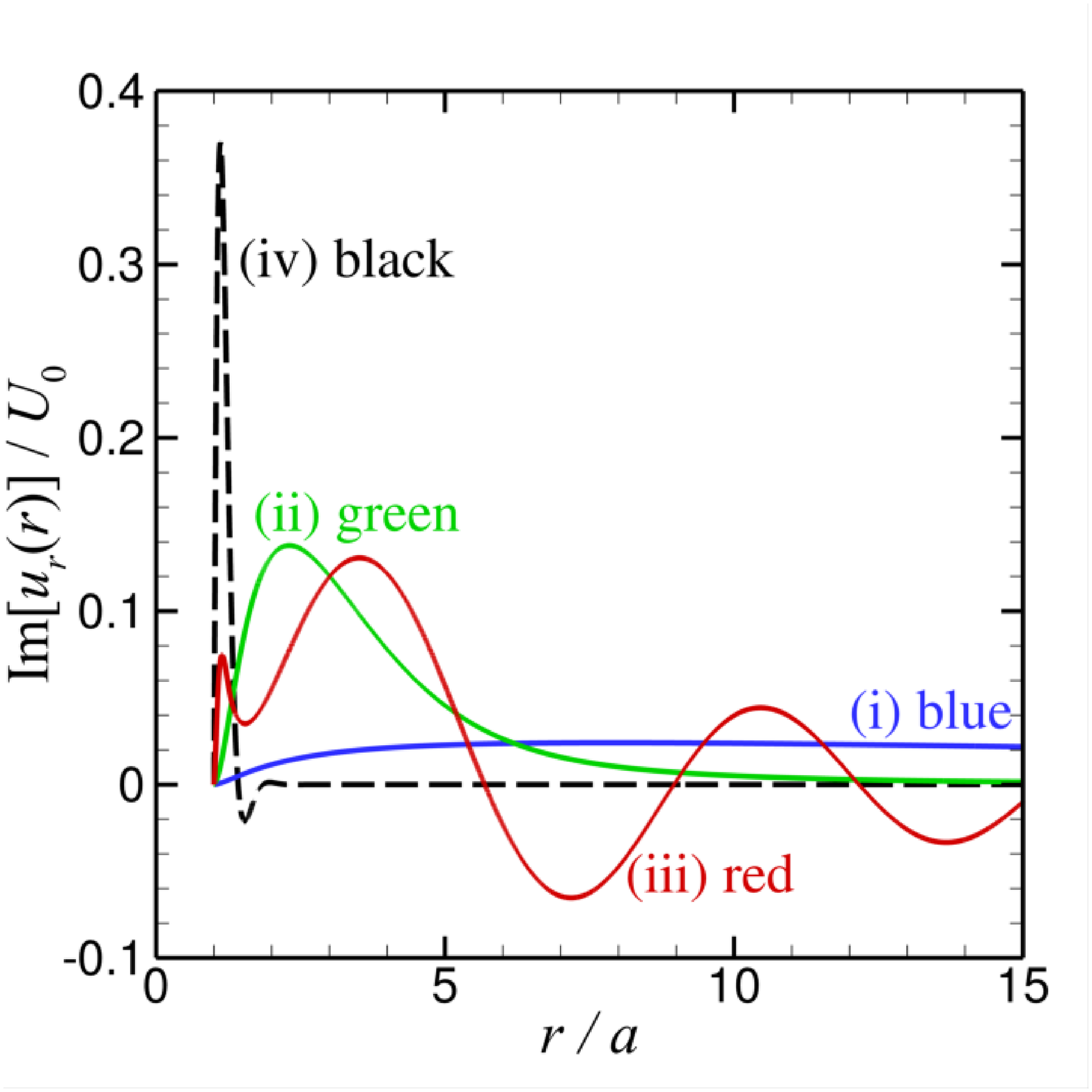}}
\subfloat[]{ \includegraphics[width=0.42\textwidth]{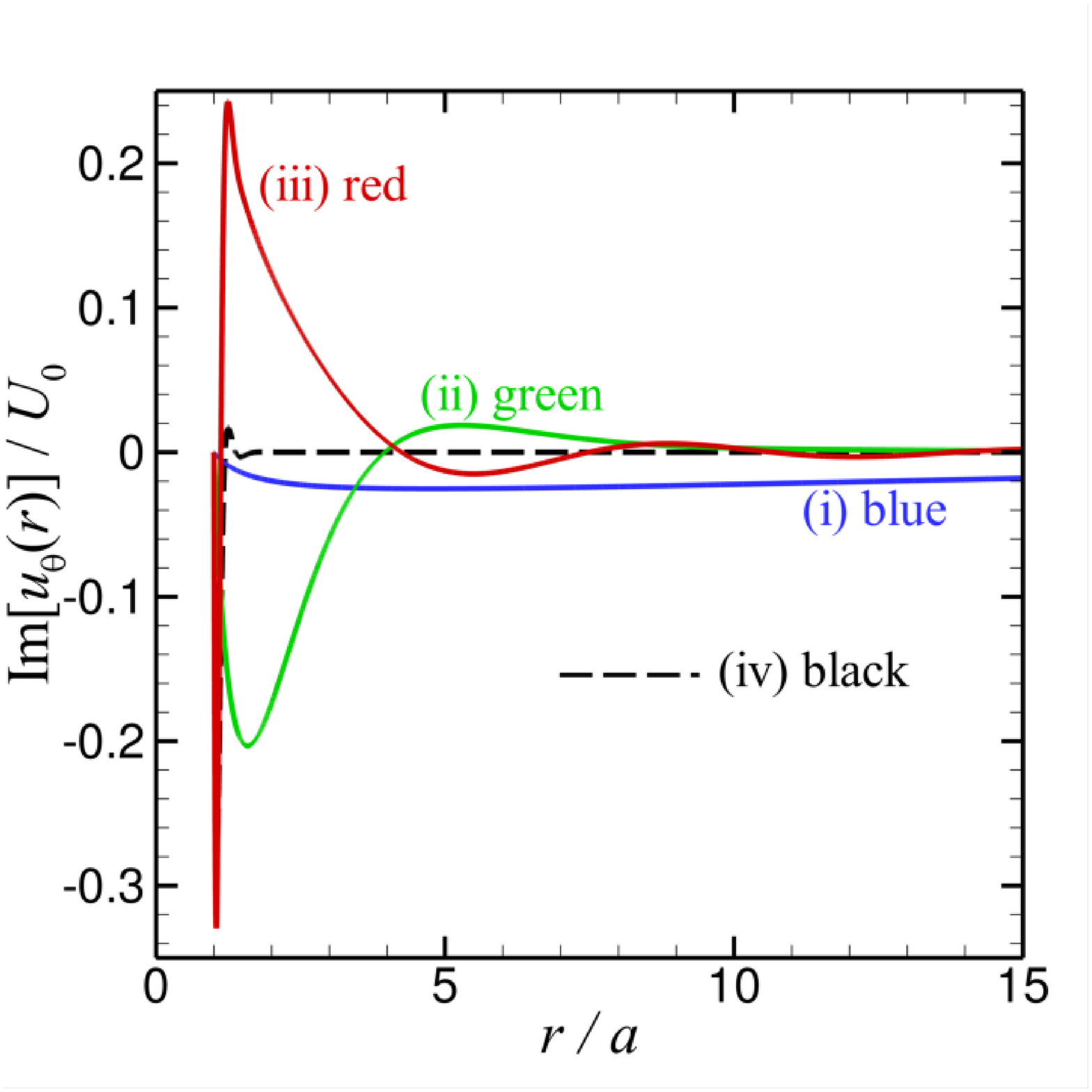}}
\caption{The primary velocity field for different values of $k_L a$ and $k_T a$ as detailed in the text for Cases (i) to (iv): (a, b)~real and (c, d)~imaginary parts of the radial, $u_r$ and tangential, $u_\theta$ primary velocity components, in which $k_T a = 0.04 + \rmi 0.04$ and $k_L a = 2.5 \times 10^{-3} + \rmi 1.35 \times 10^{-5}$ for Case (i), $k_T a = 0.7 + \rmi 0.7$ and $k_L a = 0.05 + \rmi 2.7 \times 10^{-4}$ for Case (ii), $k_T a = 14 + \rmi 14$ and $k_L a = 1.0  + \rmi 5.4 \times 10^{-3}$ for Case (iii) and $k_T a = 14 + \rmi 14$ and $k_L a=7.4 + \rmi 5.9$ for Case (iv).
}  \label{Fig:velocities}
\end{figure*}

Some representative results will be shown next. In Fig.~\ref{Fig:velocities}(a-d), we present the real and imaginary parts of the radial and tangential components of the primary velocity from the surface of the sphere at $r/a = 1$ to the far field at $r/a =15$. Values of the parameters, $k_T a$ and $k_L a$ are chosen to reflect the main differences in physical behaviour which are expected to occur if $|k_T a|$ and $|k_L a|$ are larger or smaller compared to unity in different combinations, but subject to the constraint that $|k_L^2| \leq (3/4)|k_T^2|$. The primary flow velocity components are scaled as $u_r/U_0$ and $u_\theta/U_0$,

Results for the primary velocity field in Fig.~\ref{Fig:velocities}(a-d) are given for the following 4 cases:
\begin{enumerate}[label=(\roman*)]
    \item both $|k_T a|$, $|k_L a| \ll 1$ with $k_T a = 0.04 + \rmi 0.04$ and $k_L a = 2.5 \times 10^{-3} + \rmi 1.35 \times 10^{-5}$ (blue curves). The real parts of $u_r$ and $u_\theta$ have large magnitudes and long range, varying monotonically with $r/a$. The imaginary parts are also monotonic with $r/a$ but the magnitudes are much smaller than the real parts.
    
    \item $|k_T a| \sim 1$ and $|k_L a| \ll 1$ with $k_T a = 0.7 + \rmi 0.7$ and $k_L a = 0.05 + \rmi 2.7 \times 10^{-4}$ (green curves). The real and imaginary parts of $u_r$ and $u_\theta$ have comparable magnitude but only the real part of $u_r$ is monotonic.
    
    \item $|k_T a| \gg 1$ and the real part of $k_L a \sim 1$ but the imaginary part of $k_L a$ is small with $k_T a = 14 + \rmi 14$ and $k_L a = 1.0  + \rmi 5.4 \times 10^{-3}$ (red curves). The real and imaginary part of $u_r$ and $u_\theta$ all change sign as $r/a$ increases from the sphere surface.
    
    \item both $|k_T a|$, $|k_L a| \gg 1$ and $k_L a$ has a large imaginary part with $k_T a = 14 + \rmi 14$ and $k_L a=7.4 + \rmi 5.9$ (black curves). In this case, viscous effects dominate and all flow is confined to within a thin boundary layer adjacent to the sphere surface.
\end{enumerate}

\begin{figure*}[t]
\centering{}
\subfloat[]{ \includegraphics[width=0.32\textwidth]{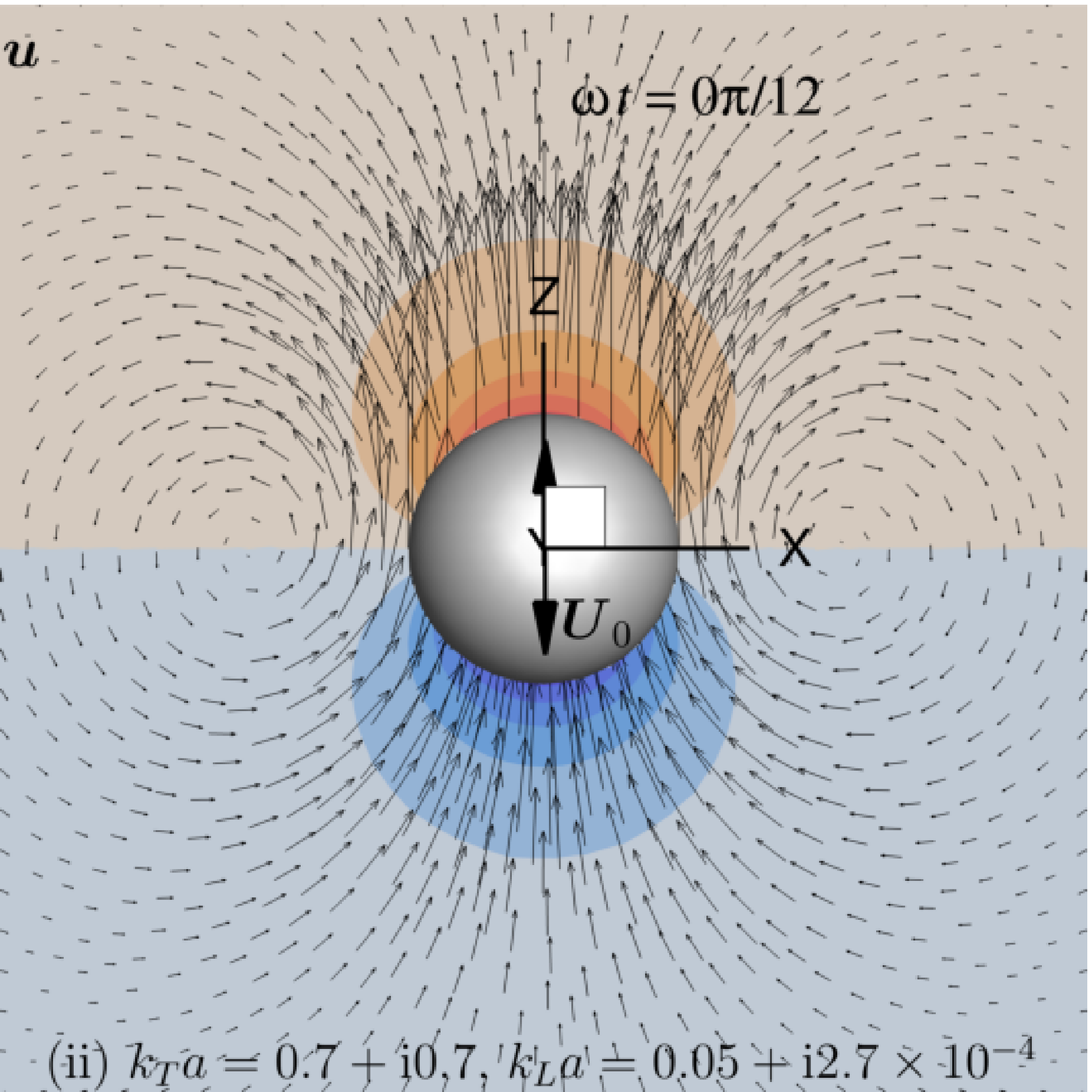}}
\subfloat[]{ \includegraphics[width=0.32\textwidth]{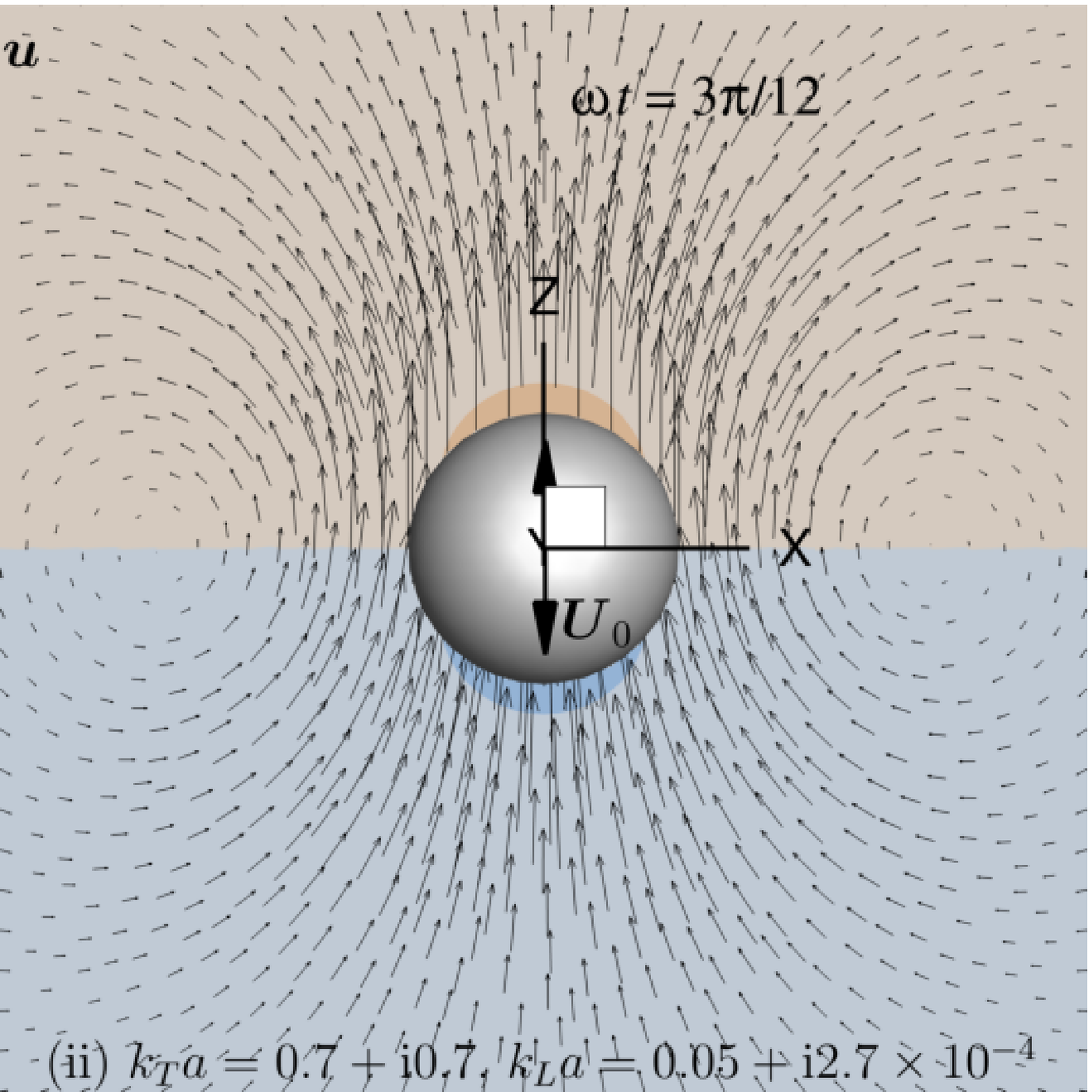}}
\subfloat[]{ \includegraphics[width=0.32\textwidth]{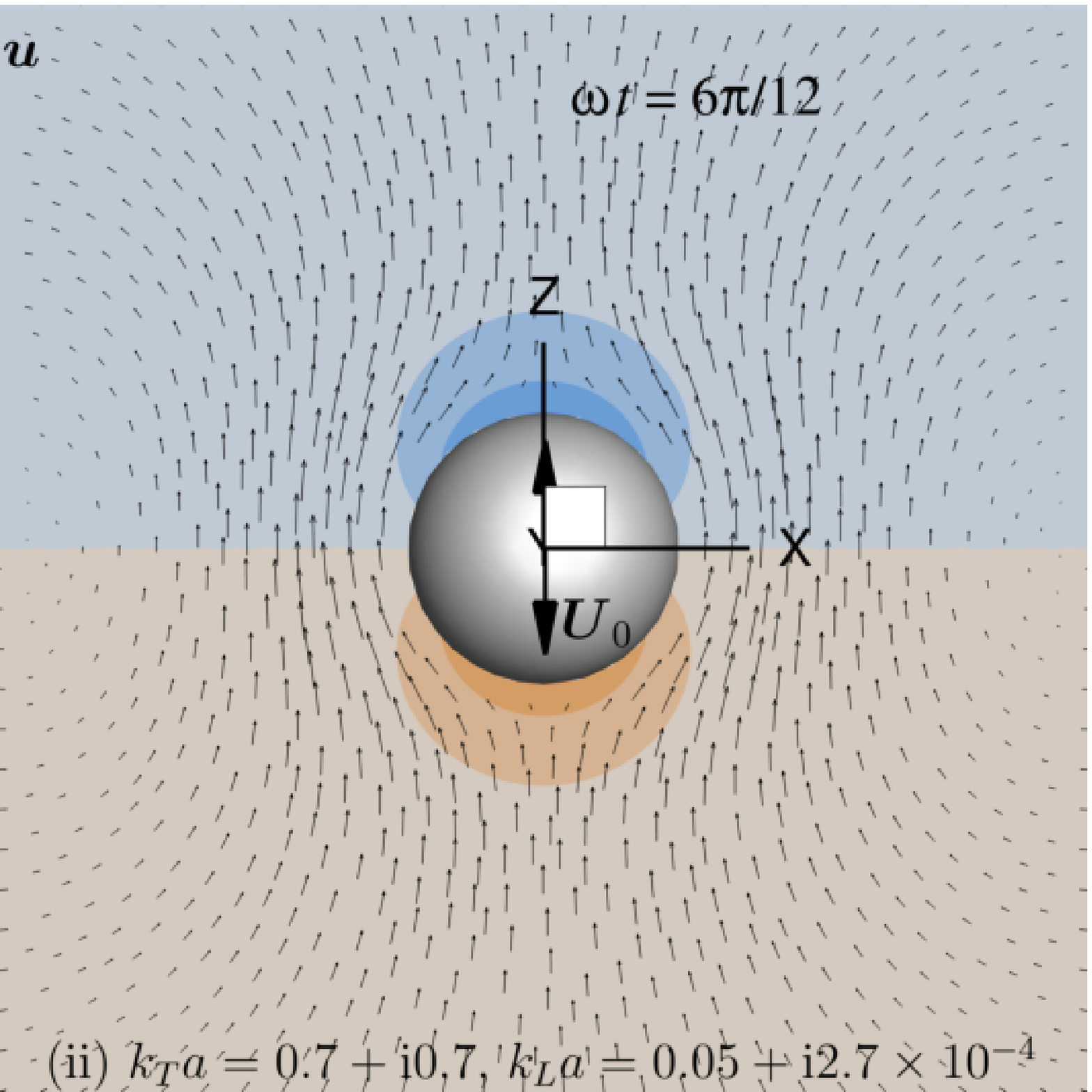}}
\caption{Primary flow velocity vector field and pressure distribution (in color) at time frame: (a) $\omega t = 0$, (b) $\omega t = 3\pi /12$, (c) $\omega t = 6\pi /12$ on $xz$ plane for Case (ii) with $k_T a = 0.7 + \rmi 0.7$ and $k_L a = 0.05 + \rmi 2.7 \times 10^{-4}$ (Multimedia view). 
}  \label{Fig:caseii}
\end{figure*}

\begin{figure*}[t]
\centering{}
\subfloat[]{ \includegraphics[width=0.32\textwidth]{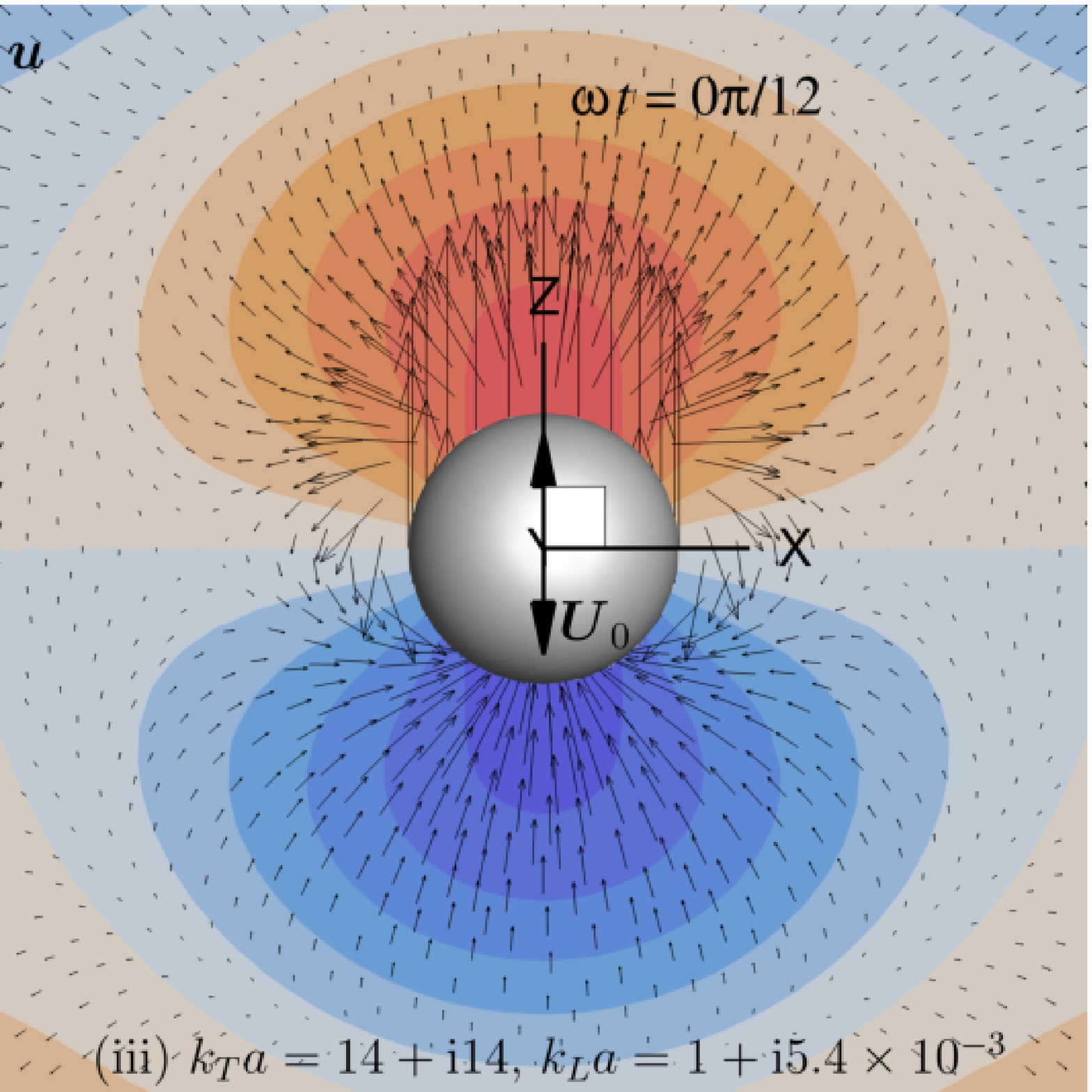}}
\subfloat[]{ \includegraphics[width=0.32\textwidth]{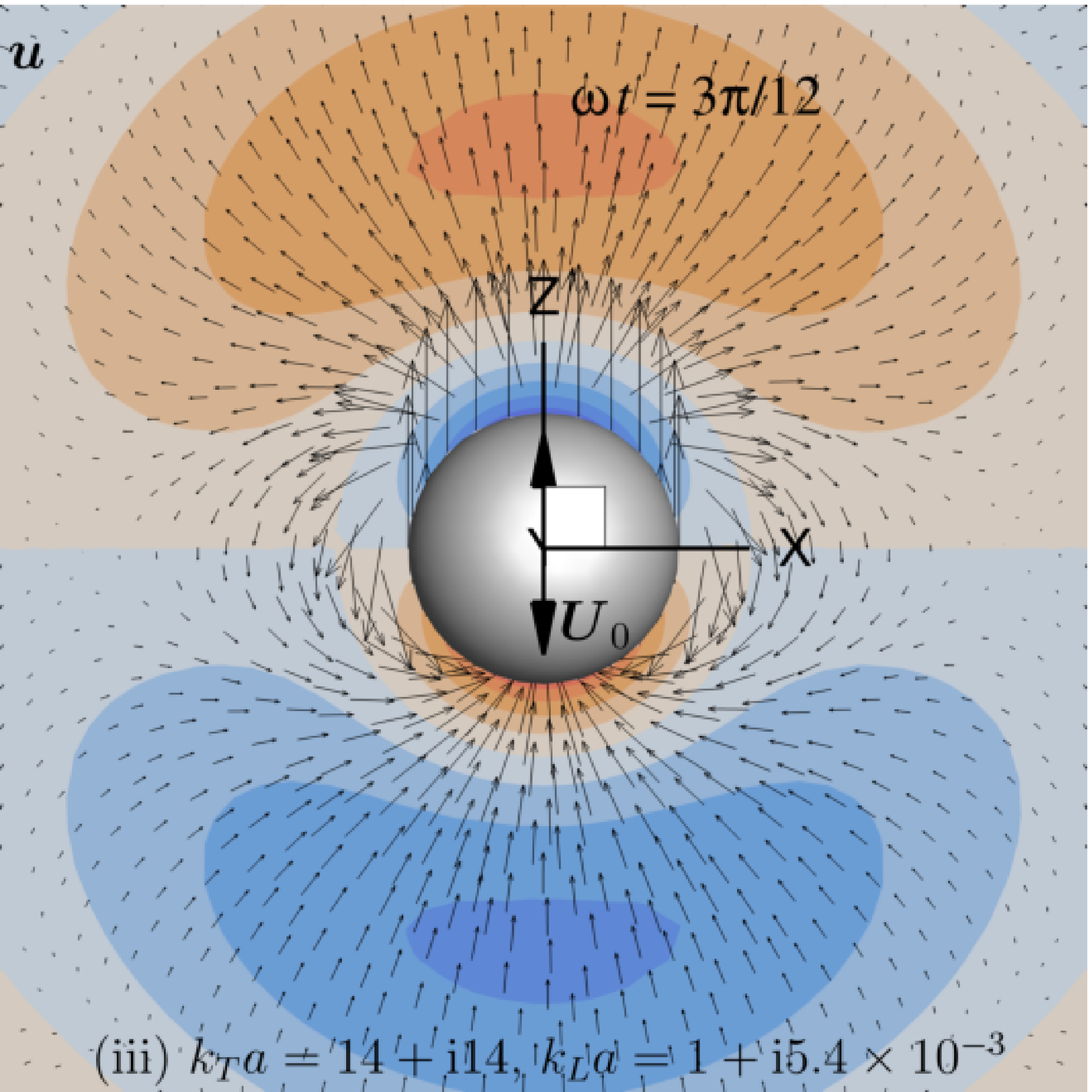}}
\subfloat[]{ \includegraphics[width=0.32\textwidth]{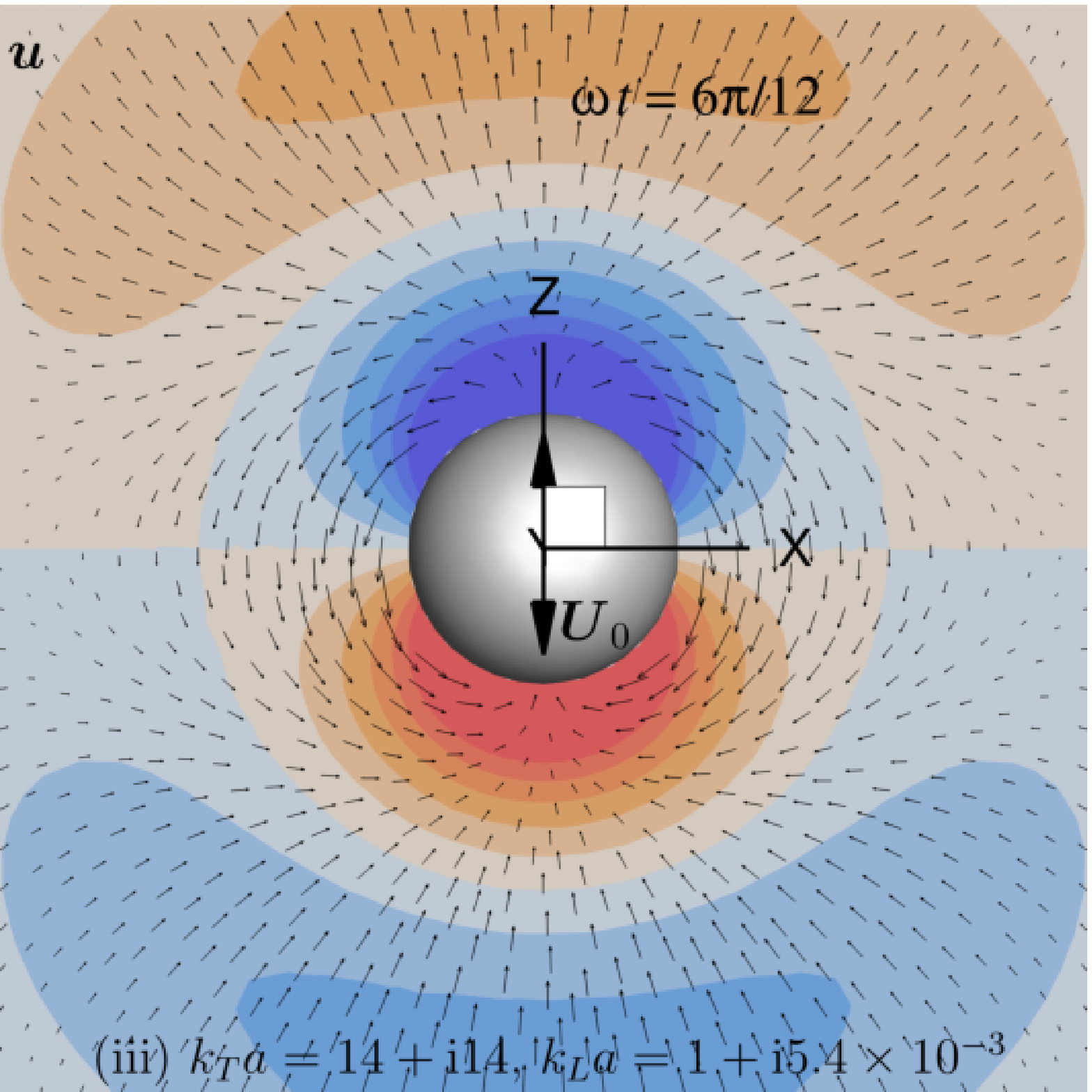}}
\caption{Primary flow velocity vector field and pressure distribution (in color) at time frame: (a) $\omega t = 0$, (b) $\omega t = 3\pi /12$, (c) $\omega t = 6\pi /12$ on $xz$ plane for Case (iii) when $k_T a = 14 + \rmi 14$ and $k_L a = 1.0  + \rmi 5.4 \times 10^{-3}$ (Multimedia view).
}  \label{Fig:caseiii}
\end{figure*}

Plots of the primary velocity, $\boldsymbol u$, (Eq.~\ref{eq:ur_utheta}) and pressure, $p$, (Eq.~\ref{eq:primary_p}), fields at selected time instants are shown in Fig.~\ref{Fig:caseii} for Case (ii) and in Fig.~\ref{Fig:caseiii} for Case (iii). These images correspond to snapshots of the animation movie that can be found in the supplementary material. Of special interest are the `vortex-alike' structures appearing to the left and right of Fig.~\ref{Fig:caseii}(a) and (b) in the primary flow patterns.

\section{Recovery of classical solutions for the primary flow}\label{sec:classicalsol}

In this section we will investigate several limits for the primary flow field, namely the small viscosity limit which will lead to potential flow, the large radius limit leading to the flat plate solution and the infinite sound speed that together with large viscosity leads to the classical Stokes flow solution.

\subsection{Small viscosity limit, thin boundary layer: potential flow}\label{sec:potential} \label{sub_sec:thin_boundary_layer}
%

If viscous effects are small, we  expect a very thin boundary layer. From (\ref{eq:kL_and_kT}), a vanishing viscosity corresponds to the limit $|k_T a|\gg|k_L a|$ and hence $e^{\rmi k_Tr}\ll1$ for $r > a$ due to the imaginary part of $k_T$. From (\ref{eq:C1_C2}) we find 
\begin{equation*}
\lim_{|k_T a| \to \infty} C_2 =\frac{1}{1+2 G(k_La)}e^{- \rmi k_La}
\end{equation*}
and (\ref{eq:ur_utheta}) becomes:
\begin{equation} \label{eq:ThinLayer1}
\begin{aligned}
\lim_{|k_T a|\to \infty} \boldsymbol u 
=  \frac{e^{\rmi k_L(r -a) }}{1+2 G(k_La)}\; \frac{a}{r} U_0 \Big[[1+2G(k_Lr)]\cos \theta \; \boldsymbol e_r + G(k_Lr) \sin \theta \; \boldsymbol e_\theta\Big].
\end{aligned}
\end{equation}
This solution is consistent with Landau and Lifshitz~\citep[p. 286, {\S}74 Problem 1]{bookLandau_FM}.

If in addition, $|k_L a| \rightarrow 0$, we recover the potential flow solution for the fluid velocity around a sphere moving at a constant velocity:
\begin{equation}
    \lim_{|k_L a|\to 0 \; ; \; |k_T a|\to \infty} \boldsymbol u 
=   \frac{a^3}{r^3} U_0 \Big[\cos \theta \; \boldsymbol e_r +\frac{1}{2} \sin \theta \; \boldsymbol e_\theta \Big].
\end{equation}
This solution is consistent with Landau and Lifshitz~\citep[p. 21--22, {\S}10 Problem 2]{bookLandau_FM}.

\begin{figure*} 
\centering{}
\subfloat[]{ \includegraphics[width=0.42\textwidth]{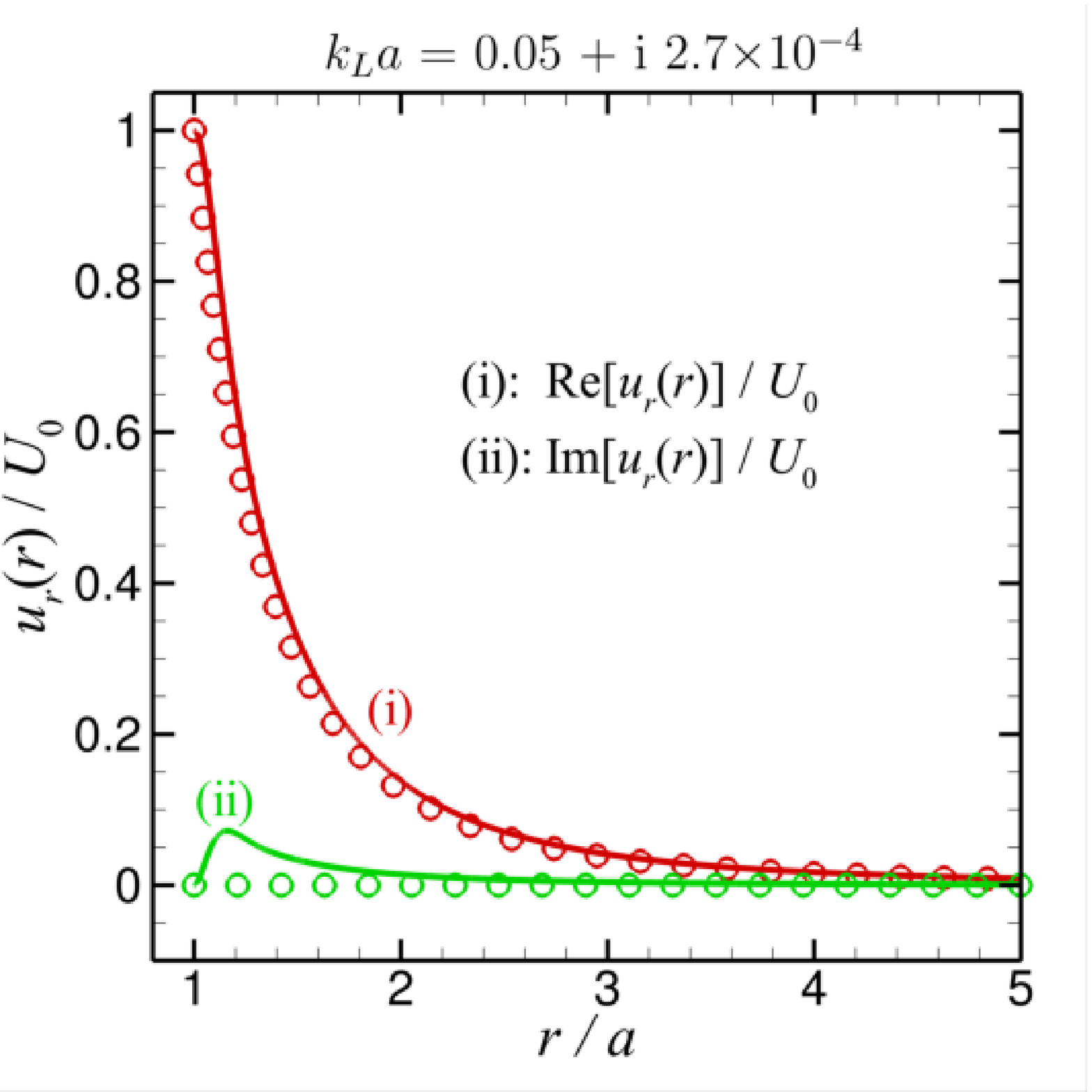}}
\subfloat[]{ \includegraphics[width=0.42\textwidth]{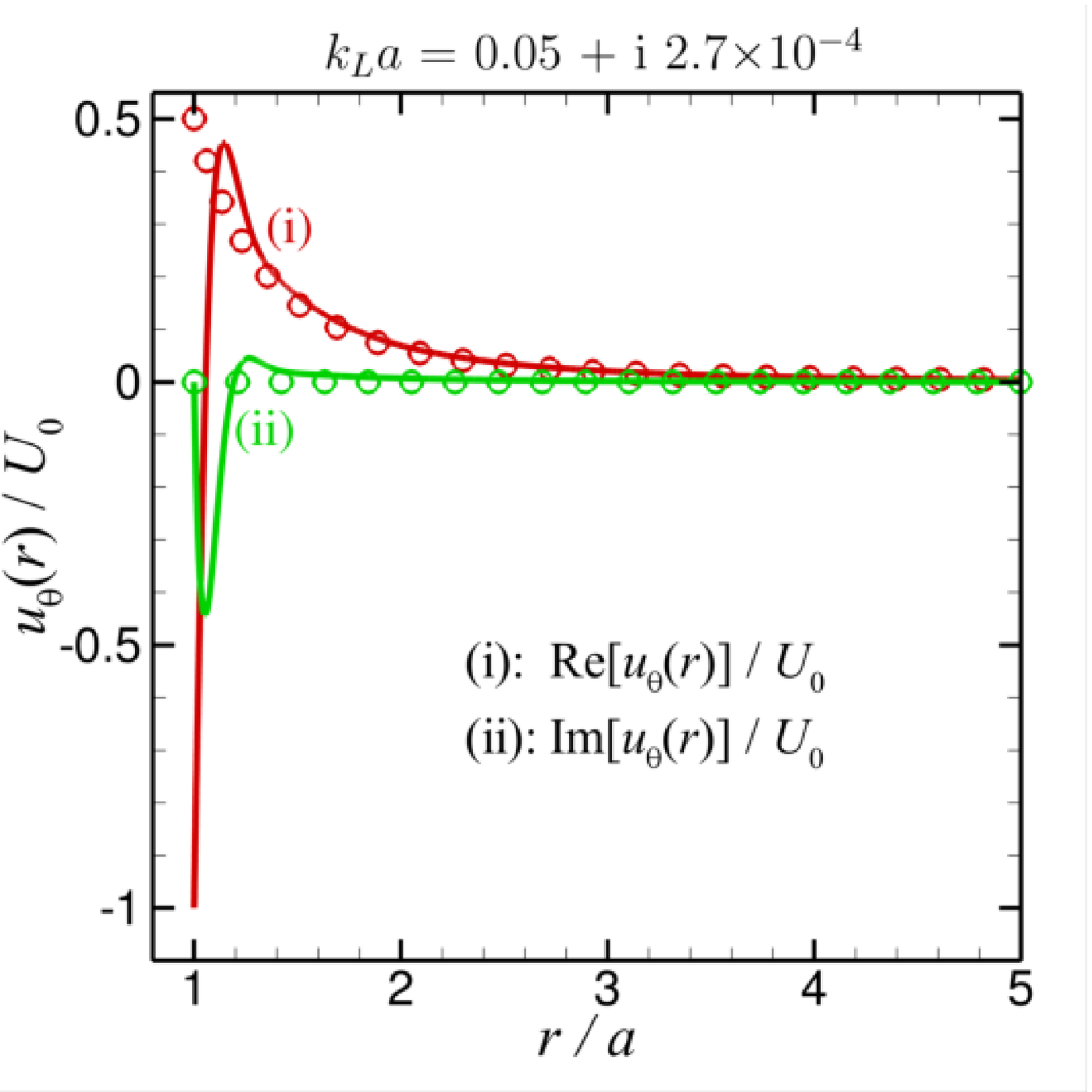}}
\caption{Comparison of the primary velocity fields at $k_Ta = 14 + \rmi \, 14$ and $k_L a = 0.05 + \rmi \, 2.7\times10^{-4}$ between the results obtained by~(\ref{eq:ur_utheta}) (lines) and those of the thin boundary layer limitation by using~(\ref{eq:ThinLayer1}) (symbols): (a) the radial, $u_r$ and (b) tangential, $u_\theta$ primary velocity components.
}  \label{Fig:potcomparison}
\end{figure*}

Fig.~\ref{Fig:potcomparison} shows the flow patterns with a carefully selected set of $k_Ta$ and $k_La$ values, $k_Ta = 14 + \rmi \, 14$ and $k_L a = 0.05 + \rmi \, 2.7\times10^{-4}$ since it is close to the thin boundary layer limit as $|k_Ta|\gg1$ and $|k_Ta| \gg |k_La|$. As demonstrated in Fig.~\ref{Fig:potcomparison}, when $|k_Ta| \gg |k_La|$, outside the thin boundary layer, the results obtained with~(\ref{eq:ur_utheta}) for the primary velocity field are in good agreement with the results using~(\ref{eq:ThinLayer1}) for the thin boundary layer potential flow limit. Also, for this particular case, as the imaginary part of $k_La$ is close to zero, the imaginary parts of the primary flow velocity components calculated by~(\ref{eq:ThinLayer1}) almost vanish, as displayed by the green circles in Fig.~\ref{Fig:potcomparison}. Note that $\rmRe [u_\theta(r)]$ closely follows the potential flow results for $r/a > 1.5$, but, in order to satisfy the no-slip condition, bends over sharply in the region $r/a=1$ to $1.5$ to satisfy the no slip condition $u_\theta(a)=-1$. In contrast, the potential flow solution is $u_\theta(a)=0.5$.

\subsection{Large radius: flat plate limit} \label{sec:flat_plate}
If the radius of the sphere, $a$, is very large, there are two locations of special interest: the front of the sphere at $\theta=0$ and the side of the sphere at $\theta = \pi/2$. 

Consider first the side at which $\cos \theta =0$ and $\sin \theta = 1$. From (\ref{eq:u_general}) we then find $\boldsymbol u = -u_\theta(r) \boldsymbol e_z$ since $\boldsymbol U_0 = -U_0 \boldsymbol e_z$. And setting $r/a \rightarrow 1$ in (\ref{eq:ur_utheta}), we find
\begin{equation}\label{eq:Flatplate1}
    \frac{u_\theta(r)}{U_0}=C_1 e^{\rmi k_Tr}[1+G(k_Tr)]+C_2e^{\rmi k_Lr} G(k_Lr).
\end{equation}
For a large sphere, $|k_L a|, |k_T a| \gg 1$, so $G(k_Lr), G(k_Tr) \rightarrow 0$ and $C_1 \rightarrow -e^{\rmi k_Ta}$ and the velocity becomes:
\begin{equation}
\label{eq:LimFlatPlate2}  
\lim_{|k_L a|, |k_T a| \to \infty \; ; \; \theta=\pi/2}\boldsymbol u =   -e^{\rmi k_T(r-a)} \boldsymbol U_0.
\end{equation}
In the time domain, this is equivalent to the well-known Stokes oscillatory boundary thickness equation \citep{Stokes1850, bookSchlichting, Ramos2001}: $\boldsymbol u= -\boldsymbol U_0 e^{-ky} \cos(\omega t - ky)$, with $k=\sqrt{\omega\rho_0/(2\mu)}$, the real part of $k_T=(1+\rmi)\sqrt{\omega\rho_0/(2\mu)}$, and $y=r-a$ being the distance from the flat surface. Thus, the solution at the side of the sphere tends towards the Stokes vibrating boundary layer theory for a flat plate when the radius of the sphere is large enough. 

For the solution in front of the sphere at $\theta =0$, in the large $a$ or flat plate limit with $\boldsymbol U_0 = U_0 \boldsymbol e_z$, the velocity in (\ref{eq:u_general}) becomes $\boldsymbol u = u_r(r) \boldsymbol e_z$ and with $a/r \rightarrow 1$, we have:
\begin{equation}\label{eq:Flatplate2} 
    \frac{u_r(r)}{U_0}=2 e^{\rmi k_Tr}C_1 G(k_Tr) +C_2e^{\rmi k_Lr} [1+2G(k_Lr)].
\end{equation}
Again with $G(k_Lr), G(k_Tr) \rightarrow 0$ for a large sphere, $|k_L a|, |k_T a| \gg 1$, we have the limiting solution
\begin{equation}
\label{eq:LimFlatPlate3}  
\lim_{|k_L a|, |k_T a|  \to \infty \; ; \; \theta=0}\boldsymbol u =   e^{\rmi k_L(r-a)} \boldsymbol U_0
\end{equation}
which represents a plane sound wave propagating out of an oscillating plate.

\subsection{Infinite sound speed: (oscillatory) incompressible Stokes flow limit
} \label{sec:steady_stream}

\begin{figure*} 
\centering{}
\subfloat[]{ \includegraphics[width=0.42\textwidth]{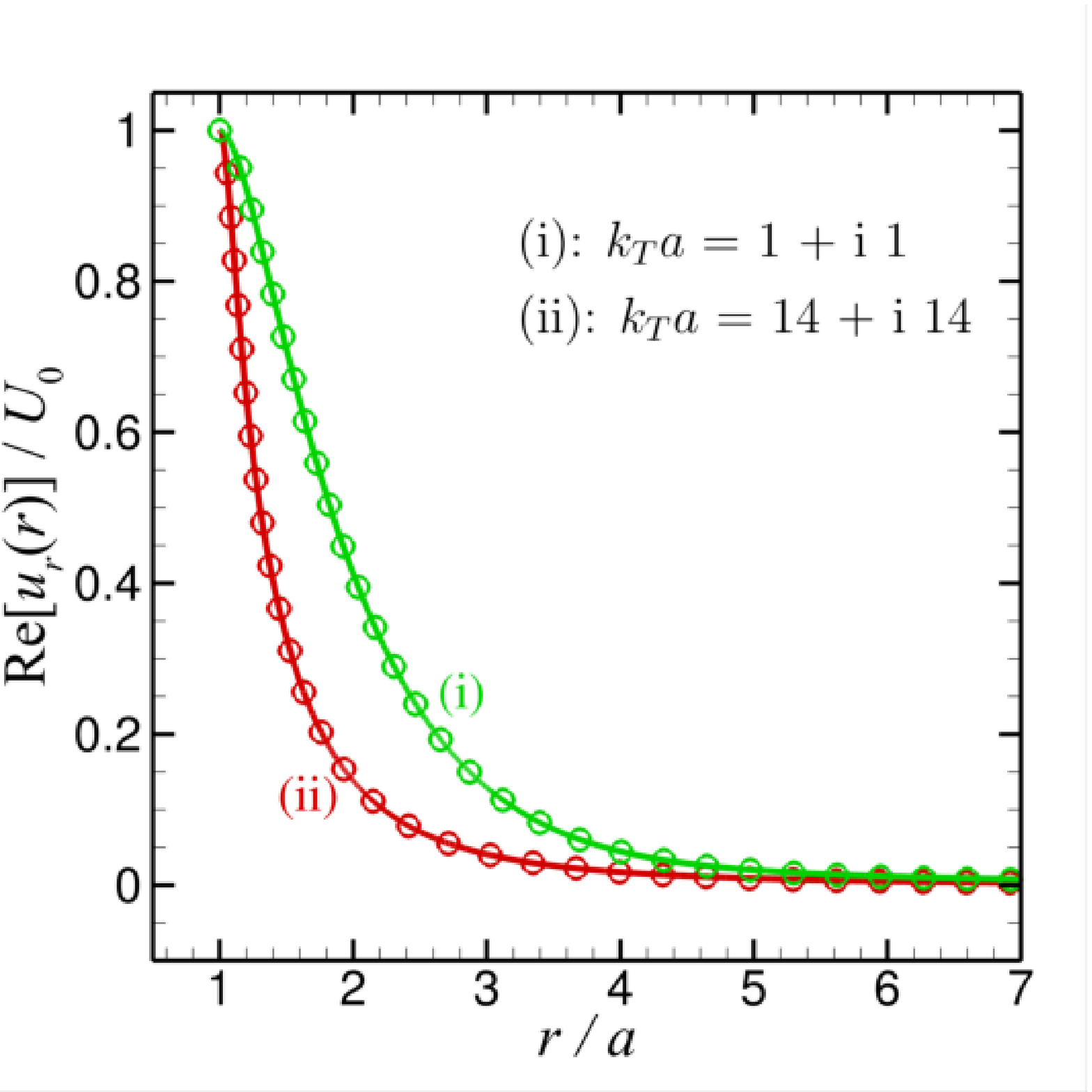}}
\subfloat[]{ \includegraphics[width=0.42\textwidth]{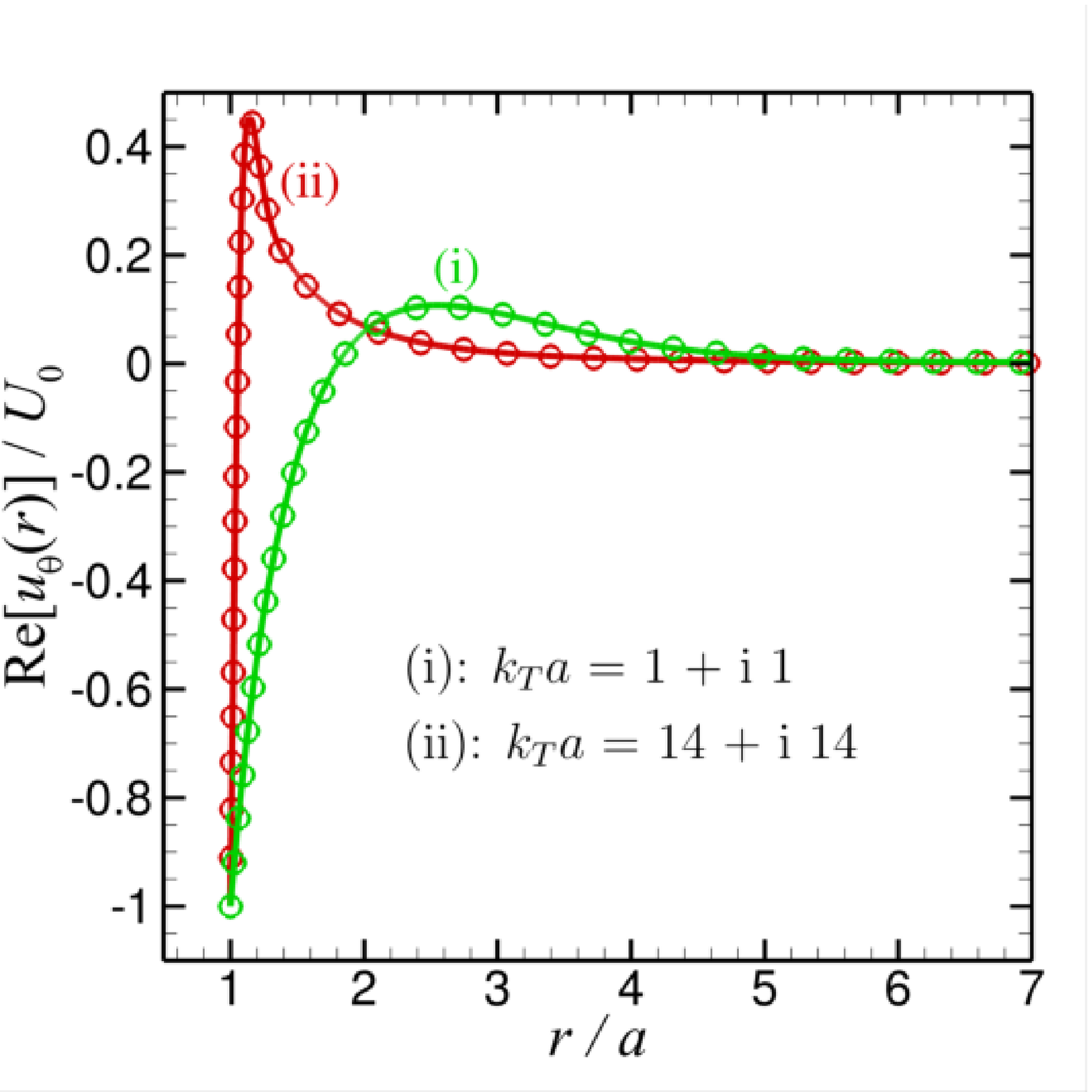}}
\\[-12pt]
\subfloat[]{ \includegraphics[width=0.42\textwidth]{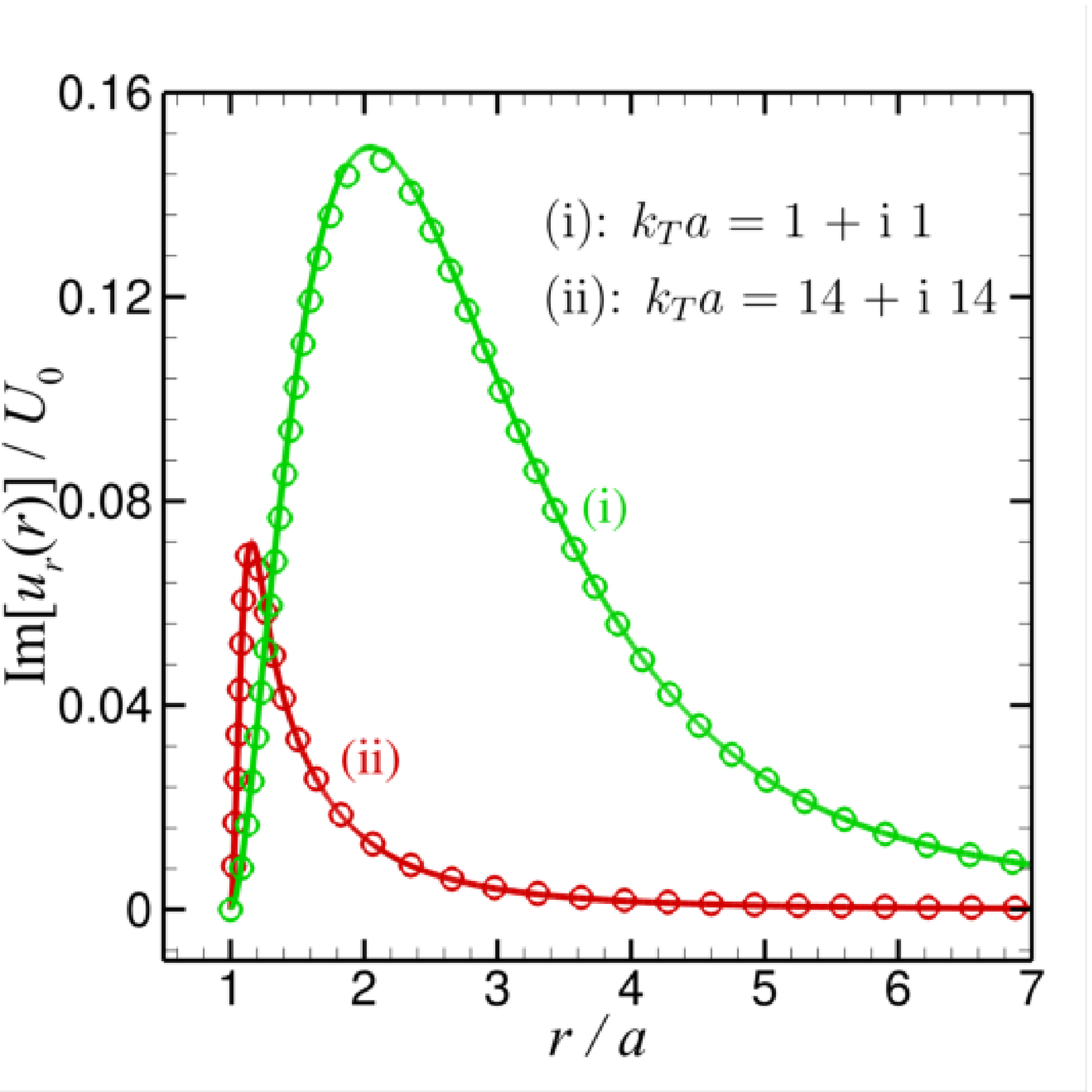}}
\subfloat[]{ \includegraphics[width=0.42\textwidth]{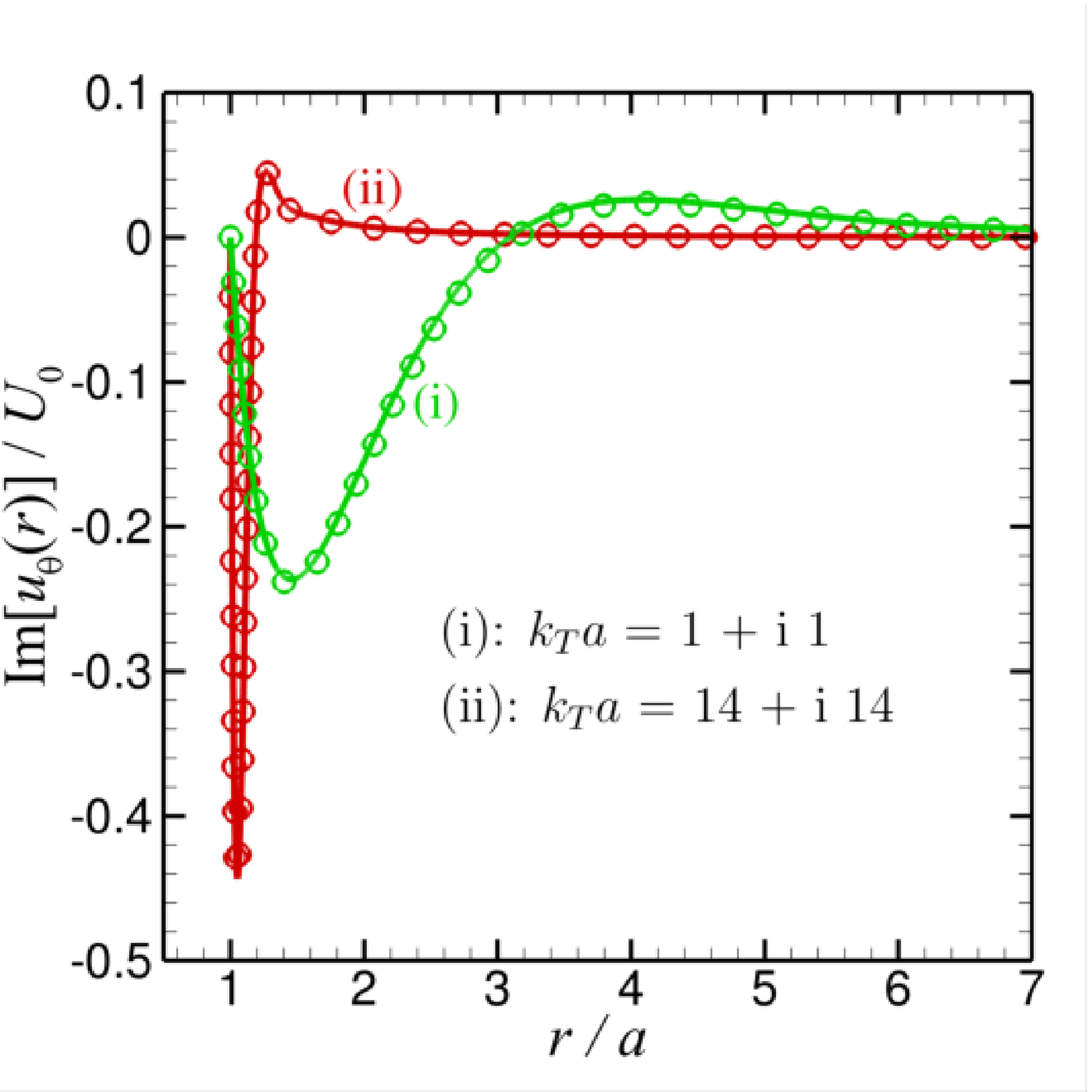}}
\caption{Comparison of the primary velocity fields at  $k_L a = 0.01 + \rmi \, 0.001$ with (i) $k_Ta = 1 + \rmi \, 1$ and (ii) $k_Ta = 14 + \rmi \, 14$ between the results obtained by Eq.~(\ref{eq:ur_utheta}) (lines) and those by using Eq.~(\ref{eq:ur_utheta_LL}) for the incompressible Stokes flow limit (symbols): (a, b)~real and (c, d)~imaginary parts of the radial, $u_r$ and tangential, $u_\theta$ primary velocity components.
}  \label{Fig:LLcomparison}
\end{figure*}

The incompressibility of the fluid implies that $c_0 \rightarrow \infty$. In this case, we can take the limit of $|k_L a| \rightarrow 0$ while keeping $|k_T a|$ finite. As such, with the help of the Maclaurin series expansion of $e^{\mathrm{i}\, k_L a}$, we have 
\begin{subequations} \label{eq:LimLL1} 
\begin{align} 
\lim_{|k_L a|\to 0} C_1 & = -\frac{3 G(k_La)}{2G(k_La)}e^{-\rmi k_T a} \rightarrow  -\frac{3}{2} e^{-\rmi k_T a},
\\
\lim_{|k_L a|\to 0} C_2 &=\frac{1+3 G(k_Ta)}{e^{ik_La}[2G(k_La)]} \rightarrow -\frac{(k_L a)^2}{2} [1+3 G(k_Ta)].
\end{align}
\end{subequations}
Introducing (\ref{eq:LimLL1}) into (\ref{eq:ur_utheta}) 
we have
\begin{subequations} \label{eq:ur_utheta_LL}
\begin{align}
    \frac{u_\theta(r)}{U_0}=& -\frac{3}{2} \; \frac{a}{r} \; \big[1+G(k_Tr)\big] \; e^{\rmi k_T (r - a)} + \frac{1+3G(k_Ta)}{2} \frac{a^3}{r^3},  \\
    \frac{u_r(r)}{U_0}=& -3 \; \frac{a}{r} \; G(k_Tr) \; e^{\rmi k_T (r - a)} + [1+3G(k_Ta)] \frac{a^3}{r^3}.
\end{align}
\end{subequations}

This solution is identical to the solution given by Landau and Lifshitz~\citep[p. 89, {\S}24 Problem 5]{bookLandau_FM}, where the velocity was written as $\boldsymbol{u} \equiv \nabla  \times \nabla \times [f_{L}(r) \boldsymbol{U}_0]$, then $u_r(r)=-(2/r) \;\rmd f_L(r)/\rmd r$ and $u_\theta(r)= (1/r) \; \rmd f_L(r)/\rmd r + \rmd^2 f_L(r)/\rmd r^2$. They showed that
 $\frac{\rmd f_{L}(r)}{\rmd r} = a_{L} \left(\frac{1}{r} - \frac{1}{\rmi k_T r^2} \right)e^{\rmi k_T r} + \frac{b_{L}}{r^2}$. This corresponds to our solution with $a_L=3\rmi a e^{-\rmi k_Ta}/(2k_T)$ and $b_L=-[1+G(k_Ta)]a^3/2$. It is also consistent with the solution of Eq. (9) from \cite{Riley1966}. As shown in Fig.~\ref{Fig:LLcomparison}, the results obtained by~(\ref{eq:ur_utheta}) for the primary velocity fields at $k_L a = 0.01 + \rmi \, 0.001$ when (i) $k_Ta = 1 + \rmi \, 1$ and (ii) $k_Ta = 14 + \rmi \, 14$ are in good agreement with the results using~(\ref{eq:ur_utheta_LL}) for the incompressible Stokes flow limit.

In order to get back the Stokes limit, we have to take the limit $|k_T a| \rightarrow 0$ as well. By using the Maclaurin series expansions of $e^{\mathrm{i}\, k_T r}$ and $e^{\mathrm{i}\, k_T a}$, we have
\begin{subequations}
\begin{align}
\lim_{|k_T a|\to0}[1+G(k_Tr)] e^{\rmi k_Tr}&=\frac{1}{2} -\frac{1}{k_T^2r^2},\\
\lim_{|k_T a|\to0}G(k_Tr) e^{\rmi k_Tr}&=-\frac{1}{2} -\frac{1}{k_T^2r^2},\\
\lim_{|k_T a|\to0}[1+3G(k_Ta)] e^{\rmi k_Ta}&=-\frac{1}{2}-\frac{3}{k_T^2a^2}.
\end{align}
\end{subequations}
Then the velocity components become: 
\begin{subequations} \label{eq:ur_utheta_LLStokes}
\begin{align}
    \frac{u_\theta(r)}{U_0} =& -\frac{3}{4} \; \frac{a}{r} +\frac{3}{2k_T^2r^2}\frac{a}{r}- \frac{1}{4}\frac{a^3}{r^3}- \frac{3}{2k_T^2 a^2} \frac{a^3}{r^3} = -\frac{3}{4} \; \frac{a}{r} - \frac{1}{4}\frac{a^3}{r^3}, \\
    \frac{u_r(r)}{U_0} =&\frac{3}{2} \; \frac{a}{r} +\frac{3}{k_T^2r^2}\frac{a}{r}- \frac{1}{2}\frac{a^3}{r^3}- \frac{3}{k_T^2 a^2} \frac{a^3}{r^3} = \frac{3}{2} \; \frac{a}{r} - \frac{1}{2}\frac{a^3}{r^3}
\end{align}
\end{subequations}
which is the velocity field for a sphere moving in Stokes flow Landau and Lifshitz~\citep[p. 50--60, \S20]{bookLandau_FM}. Note that in the above results, terms with $1/k_T^2$ cancel each other exactly out.

\section{Conclusion}\label{sec:conclusion}
The analytical solution of the flow field around a rigid sphere executing small amplitude rectilinear motion in an compressible fluid was investigated. Both the primary flow, where second order inertial effects were neglected, and the secondary flow were studied. The mathematical form of the equation that governs the primary velocity field is identical to that for the propagation of elastic waves in solids. This allows us to draw on earlier work~\cite{KlaseboerJElas2018} to obtain analytic solutions. The primary flow field was shown to adhere to all the classical analytical expressions in the small or large viscosity and/or radius limits and is valid for very thin all the way to very thick boundary layers.  

The equation that governs the (secondary) streaming flow is analogous to the problem of the flow phenomenon associated with the electrophoretic mobility of a spherical charged particle~\citep{Overbeek1941, JAYARAMAN2019845} that enables the streaming velocity field to be expressed readily in terms of the body force and the vorticity.

To the best knowledge of the authors, this is one of the few analytic results in the theory of acoustic boundary layer flow and can serve as benchmark of numerical solution schemes for more complex problems.

\begin{acknowledgments}
This work is supported in part by a Discovery Project Grant (DP170100376) to DYCC. QS is supported by a Discovery Early Career Researcher Award (DE150100169) and a Centre of Excellence Grant (CE140100003) funded by the Australian Research Council.
\end{acknowledgments}

\section*{Data Availability}
The data that support the findings of this study are available from the corresponding author upon reasonable request.

\appendix

\section{Derivation of the secondary flow field} \label{app:SecondaryFlowDerivation}

In this appendix the theoretical solution for the secondary flow field is derived as outlined in Section~\ref{sec:SolutionOutline}.

\subsection{Symmetry of the streaming equation} \label{sub_sec:formal_streaming}
From the symmetry condition of the primary velocity field in (\ref{eq:u_general}), it follows from (\ref{eq:streaming_eqn}) that the body force per unit volume that drives the streaming flow must be of the form 
\begin{equation}   \label{eq:streaming_F}
\begin{aligned}
    \boldsymbol {\mathcal{F}}(\boldsymbol x) =& \frac{\rho_0}{4}\Big[ ({\boldsymbol  u} \cdot \nabla)  {\boldsymbol  u}^{*} + ({\boldsymbol  u}^{*} \cdot \nabla)  {\boldsymbol  u} + (\nabla \cdot  {\boldsymbol  u^{*}})  {\boldsymbol  u} + (\nabla \cdot  {\boldsymbol  u})  {\boldsymbol  u}^* \Big] \\
    =&  [ F_{0}(r) + F_{2}(r) \; (3\cos^2\theta -1) ] \; \boldsymbol{e}_r + F_{1}(r) \cos\theta \; \sin\theta \; \boldsymbol{e}_\theta
\end{aligned}
\end{equation}
where the superscript `$^*$' denotes the complex conjugate. Explicit expressions for the functions $F_2(r)$, $F_1(r)$ and $F_0(r)$ in (\ref{eq:streaming_F}) for $\boldsymbol{\mathcal{F}}$ are
\begin{subequations}
\begin{align}
F_2(r) =&\frac{\rho_0}{6}\frac{\rmd u_r u_r^*}{\rmd r}+\frac{\rho_0}{12r}[2u_\theta u_\theta^*+3u_\theta u_r^*+3u_\theta^* u_r+4u_r u_r^*],  \\
F_1(r) =&\frac{\rho_0}{4}\frac{\rmd}{\rmd r}[u_\theta u_r^*+u_\theta^* u_r] +\frac{3\rho_0}{4r}(2u_\theta u_\theta^*+u_\theta u_r^*+u_\theta^*u_r), \\[3pt]
F_0(r) =& \frac{\rho_0}{6}\frac{\rmd u_r u_r^*}{\rmd r}+\frac{\rho_0}{12r}[4u_r u_r^* - 4u_\theta u_\theta^*].
\end{align}
\end{subequations}
Using this general form for $\boldsymbol {\mathcal{F}}(\boldsymbol x)$ in (\ref{eq:streaming_eqn}), the streaming velocity must have the following angular and radial dependency
\begin{equation}
   \label{eq:streaming_U} 
\begin{aligned}
    \boldsymbol{U}(\boldsymbol x) 
     = &  U_r(r) \; (3\cos^2\theta -1) \; \boldsymbol{e}_r + U_\theta(r) \cos\theta \; \sin\theta \; \boldsymbol{e}_\theta \\
     = &  \frac{S(r)}{r} \; (3\cos^2\theta -1) \; \boldsymbol{e}_r -  \frac{1}{r}\frac{\rmd}{\rmd r} \Big[r S(r)\Big] \cos\theta \; \sin\theta \; \boldsymbol{e}_\theta .
\end{aligned}
\end{equation}
The condition $\nabla \cdot \boldsymbol U(\boldsymbol x) = 0$ (see Nyborg~\cite{NyborgJASA1953} and Lee and Wang~\cite{Lee1990}) ensures that the streaming velocity is determined by the single function $S(r)$ in (\ref{eq:streaming_U}). The form for the pressure can be inferred from (\ref{eq:streaming_eqn})
\begin{equation} \label{eq:streaming_P}
    P(\boldsymbol x) =  
     P_0(r) + P_2(r) (3 \cos^2\theta - 1).
\end{equation}

Instead of solving for the three functions $S(r)$, $P_0(r)$ and $P_2(r)$, it is more convenient to work in terms of the vorticity field, $\boldsymbol W = \nabla \times \boldsymbol U$ to eliminate the pressure, $P(\boldsymbol x)$. We hereby follow very closely the theory used for electrophoresis of a moving sphere, in particular that of the PhD thesis of Overbeek~\cite{Overbeek1941} and although the theory is not identical, many of the mathematical concepts, such as vector manipulation in spherical coordinates and double integration techniques can still be utilised here. Taking the curl of (\ref{eq:streaming_eqn}) gives $\nabla \times \boldsymbol {\mathcal{F}} = \mu\nabla^2 \boldsymbol W $. Since both the body force, $\boldsymbol {\mathcal{F}}$ and the streaming velocity, $\boldsymbol U$ are independent of the azimuthal angle $\varphi$ and have no azimuthal $\varphi$-component, their curl will only have a non-zero component in the $\varphi$-direction along the unit vector, $\boldsymbol{e}_\varphi$. This then ensures that $\nabla^2 \boldsymbol W$ also points in the $\varphi$ direction. From (\ref{eq:streaming_F}) and (\ref{eq:streaming_U}) they are characterised by two functions, $f(r)$ and $W(r)$, that are only functions of the radial distance, $r$ from the centre of the sphere:
\begin{equation}  \label{eq:curl_F}
    \nabla \times \boldsymbol {\mathcal{F}}(\boldsymbol x) =   f(r) \cos\theta \; \sin\theta \; \boldsymbol{e}_\varphi,
\end{equation}
\begin{equation} \label{eq:curl_U}
    \boldsymbol{W}(\boldsymbol x) \equiv \nabla \times \boldsymbol{U}(\boldsymbol x) =  W(r) \cos\theta \; \sin\theta \; \boldsymbol{e}_\varphi
\end{equation}
with $f(r)$ that characterises the body force in (\ref{eq:curl_F}) given by
\begin{equation} 
\begin{aligned}
    f(r) =  \frac{1}{r} \left\{ \frac{\rmd}{ \rmd r} [r F_{1}(r)] + 6F_2(r) \right\}.
\end{aligned}
\end{equation}  
This completes the framework of the axisymmetric streaming flow around a sphere.

\subsection{Formal solution of the streaming equation} \label{sub_sec:formal_streaming_solution}

The method of solution involves deriving an ordinary differential equation for $W(r)$ defined in (\ref{eq:curl_U}) in terms of $f(r)$ defined in (\ref{eq:curl_F}). Then the velocity function, $S(r)$ given by (\ref{eq:streaming_U}) can be expressed in terms of $W(r)$ to give the solution for the velocity field. 

Expressing $\nabla^2 \boldsymbol{W}(\boldsymbol x)$ in spherical polar coordinates, the curl of (\ref{eq:streaming_eqn}) becomes
\begin{equation}
     f(r) = \mu r\frac{\rmd}{\rmd r}\left[\frac{1}{r^4}\frac{\rmd}{\rmd r}\Big(r^3 W(r) \Big) \right]
\end{equation}
that can be integrated immediately to give the vorticity function, $W(r)$, noting that $W(r) \rightarrow 0$ as $r \rightarrow \infty$,
\begin{equation}\label{eq:W}
    W(r) =\frac{c_w}{r^3}+W_f(r) \qquad \text{with} \qquad W_f(r) \equiv \frac{1}{\mu} \frac{1}{r^3}\int_r^\infty y^4 \left[ \int_y^\infty \frac{f(x)}{x}\rmd x \right] \rmd y.
\end{equation}
The integration constant, $c_w$ from the homogeneous solution can be determined from the boundary conditions at the sphere surface, $r=a$.

In a similar way, the streaming velocity function, $S(r)$ can be found by combining (\ref{eq:streaming_U}) and (\ref{eq:curl_U}) to give
\begin{equation} \label{eq:W_S}
    \begin{aligned}
    W(r)=-r\frac{\rmd}{\rmd r}\left[\frac{1}{r^4} \frac{\rmd}{\rmd r}\left[ r^3 S(r) \right] \right] 
    \end{aligned}
\end{equation}
that can be integrated to give $S(r)$, noting that $S(r) \rightarrow 0$ as $r \rightarrow \infty$,
\begin{equation}\label{eq:S}
    S(r) = \frac{c_s}{r^3} + \frac{c_w}{6r} - \frac{1}{r^3}\int_r^\infty y^4 \left[ \int_y^\infty \frac{W_{f}(x)}{x}\rmd x \right] \rmd y.
\end{equation}
If the boundary condition of the general problem is that the fluid velocity is prescribed on the sphere surface, this will be satisfied by the primary flow so that the boundary condition of the streaming velocity is $\boldsymbol{U}(r=a) = \boldsymbol{0}$.
Using (\ref{eq:streaming_U}), this gives $S(a) = 0$ and $\rmd [rS(r)]/ \rmd r = 0$ at $r = a$. From (\ref{eq:S}), we obtain
\begin{equation} \label{eq:cwcs}
\begin{aligned} \nonumber
  c_w =- 3 a^2\int^{\infty}_{a} \frac{W_f(x)}{x} \rmd x,\qquad 
  c_s = \frac{a^5}{2} \int^{\infty}_{a} \frac{W_f(x)}{x} 
  \rmd x + \int^\infty_a y^4 \int^\infty_y \frac{W_f(x)}{x}\rmd x \rmd y.
\end{aligned}
\end{equation}
\begin{figure*}[ht]
\centering{}
\subfloat[]{ \includegraphics[width=0.41\textwidth]{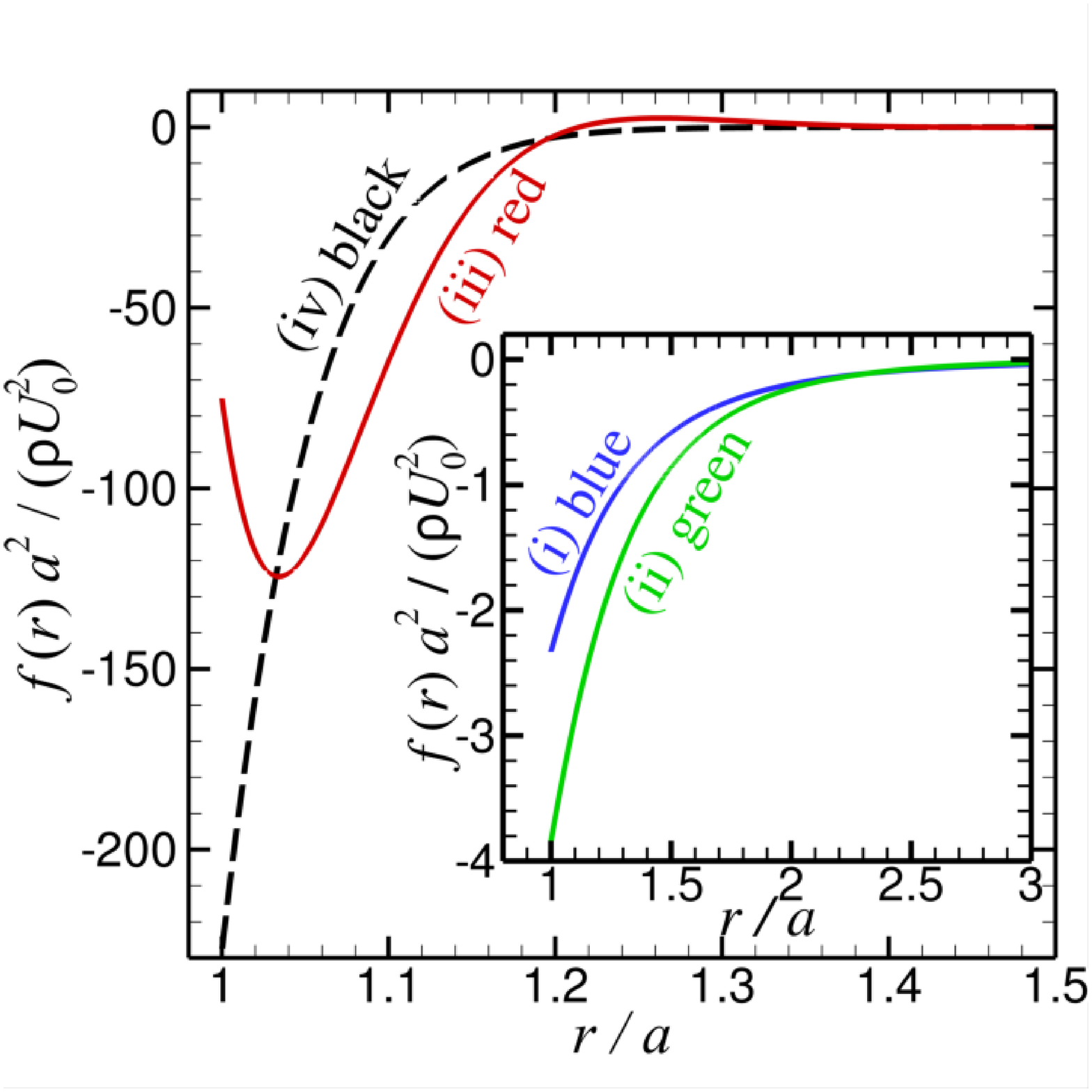}}
\subfloat[]{ \includegraphics[width=0.41\textwidth]{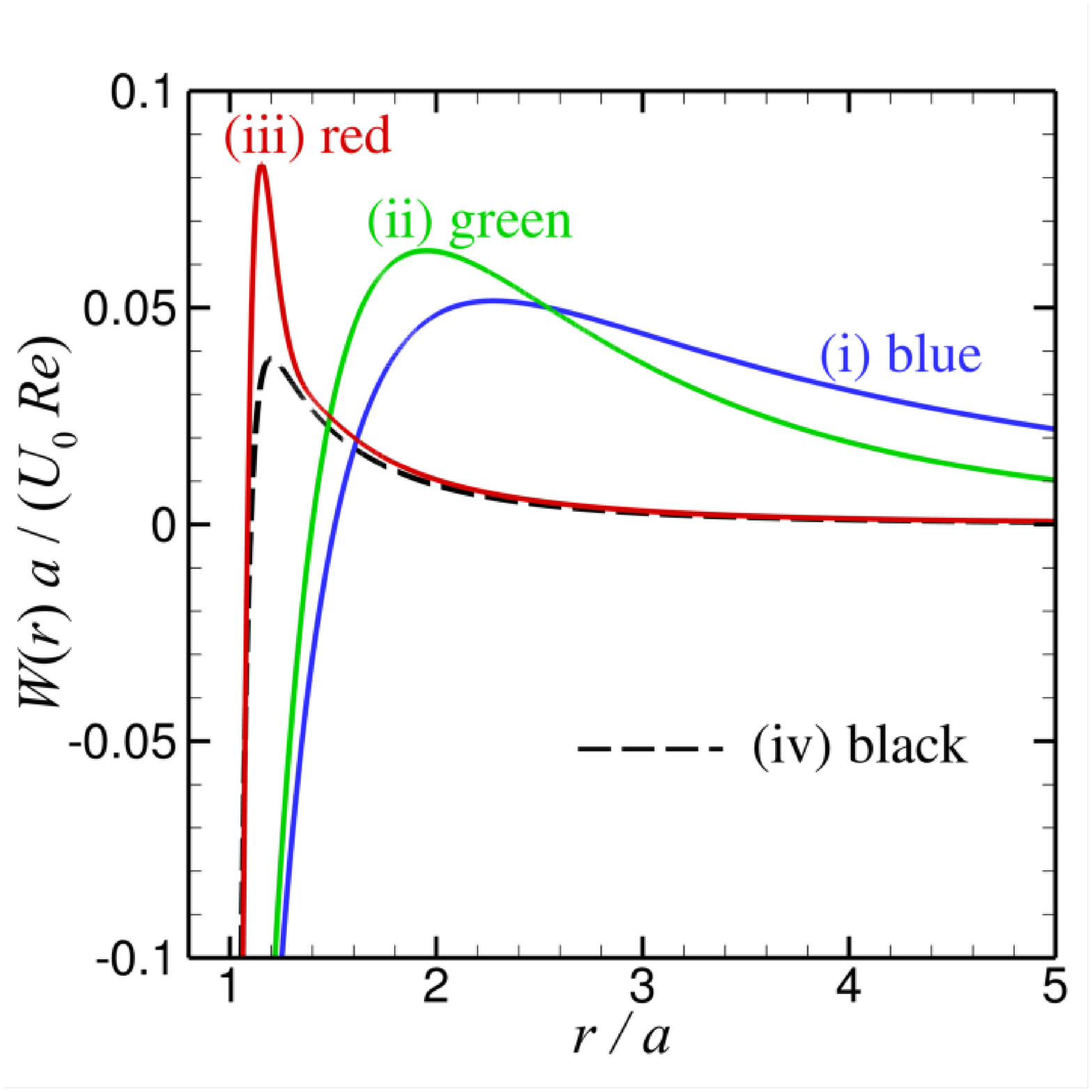}} \\ \vspace{-12pt}
\subfloat[]{ \includegraphics[width=0.41\textwidth]{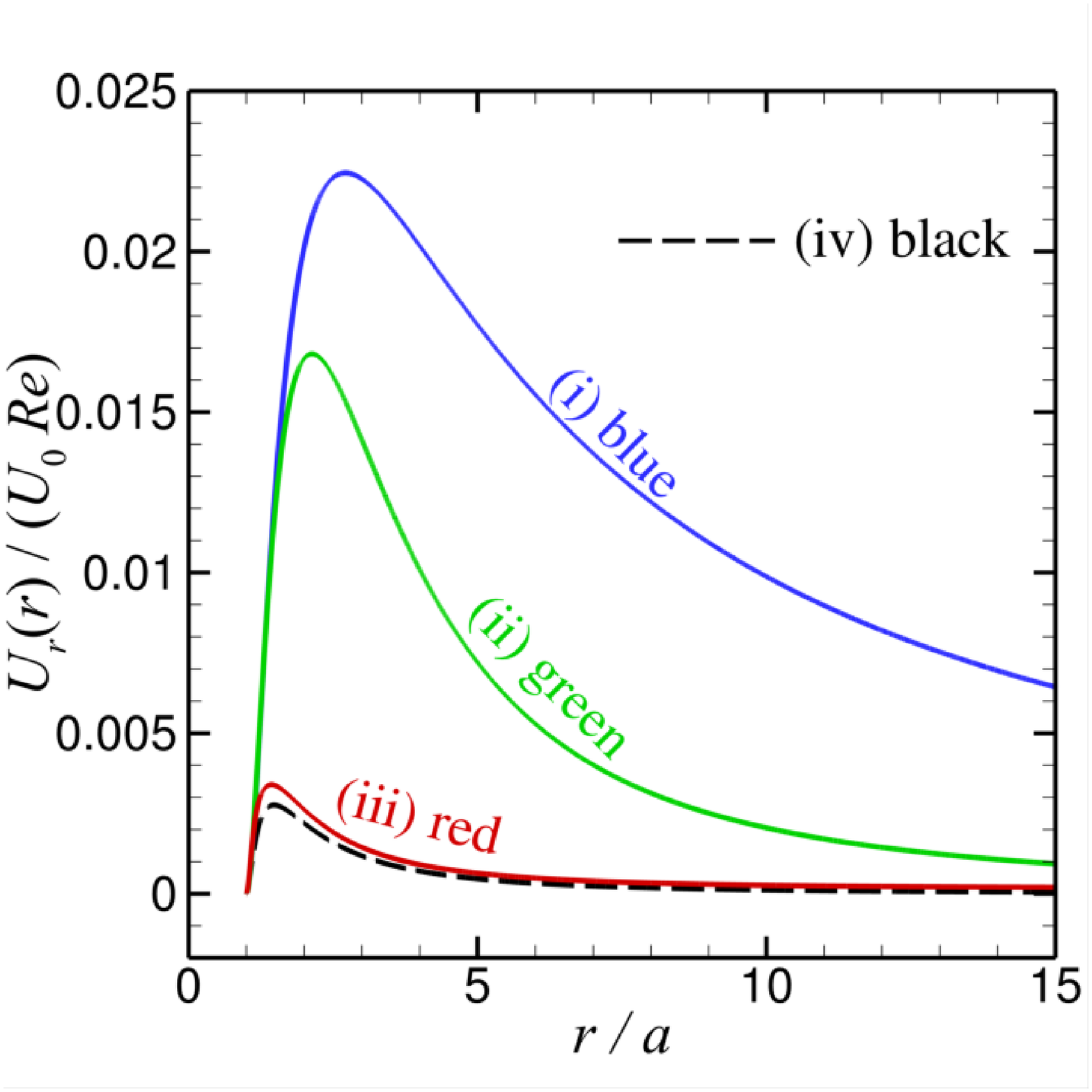}}
\subfloat[]{ \includegraphics[width=0.41\textwidth]{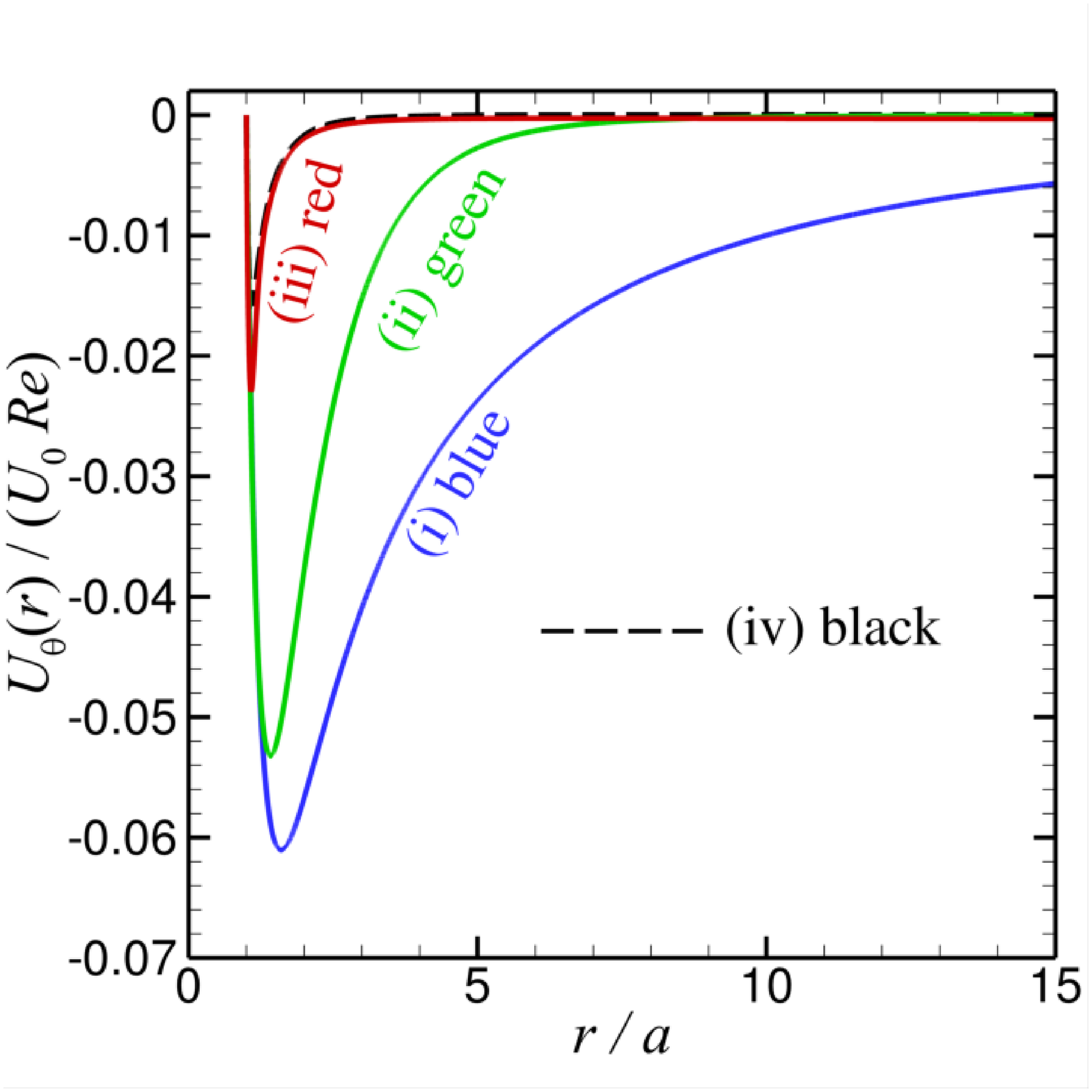}}
\caption{(a) The function $f(r)$ that characterises the curl of the body force in (\ref{eq:curl_F}) that drives streaming flow and (b) the streaming vorticity fields, $W(r)$, (c)~the radial, $U_r$ and (d) tangential, $U_\theta$ components of the streaming velocity for different values of $k_L a$ and $k_T a$ for Case (i) with $k_T a = 0.04 + \rmi 0.04$ and $k_L a = 2.5 \times 10^{-3} + \rmi 1.35 \times 10^{-5}$, Case (ii) with $k_T a = 0.7 + \rmi 0.7$ and $k_L a = 0.05 + \rmi 2.7 \times 10^{-4}$, Case (iii) with $k_T a = 14 + \rmi 14$ and $k_L a = 1.0  + \rmi 5.4 \times 10^{-3}$ and Case (iv) with $k_T a = 14 + \rmi 14$ and $k_L a=7.4 + \rmi 5.9$.
}  \label{Fig:vorticity}
\end{figure*}

The pressure functions, $P_{0}(r)$ and $P_{2}(r)$ of the secondary flow in (\ref{eq:streaming_P}) can be readily obtained: $P_{0}(r)$ can be found by taking the divergence of (\ref{eq:streaming_eqn}) to eliminate the velocity $\boldsymbol{U}(\boldsymbol{x})$ and  $P_{2}(r)$ can be obtained from the $\boldsymbol{e}_{\theta}$ component of (\ref{eq:streaming_eqn}) to give
\begin{equation}\label{eq:stream_p1}
   P_{0}(r) = -\frac{1}{2}\int_{r}^{\infty}F_{0}(r) \rmd r, \qquad P_{2}(r) = \frac{\mu r}{6}  \left[ \frac{\rmd W(r)}{\rmd r} + \frac{W(r)}{r} \right] - \frac{r}{6} F_{1} (r).
\end{equation}
\begin{figure*}[!ht]
\centering{}
\subfloat[]{ \includegraphics[width=0.42\textwidth]{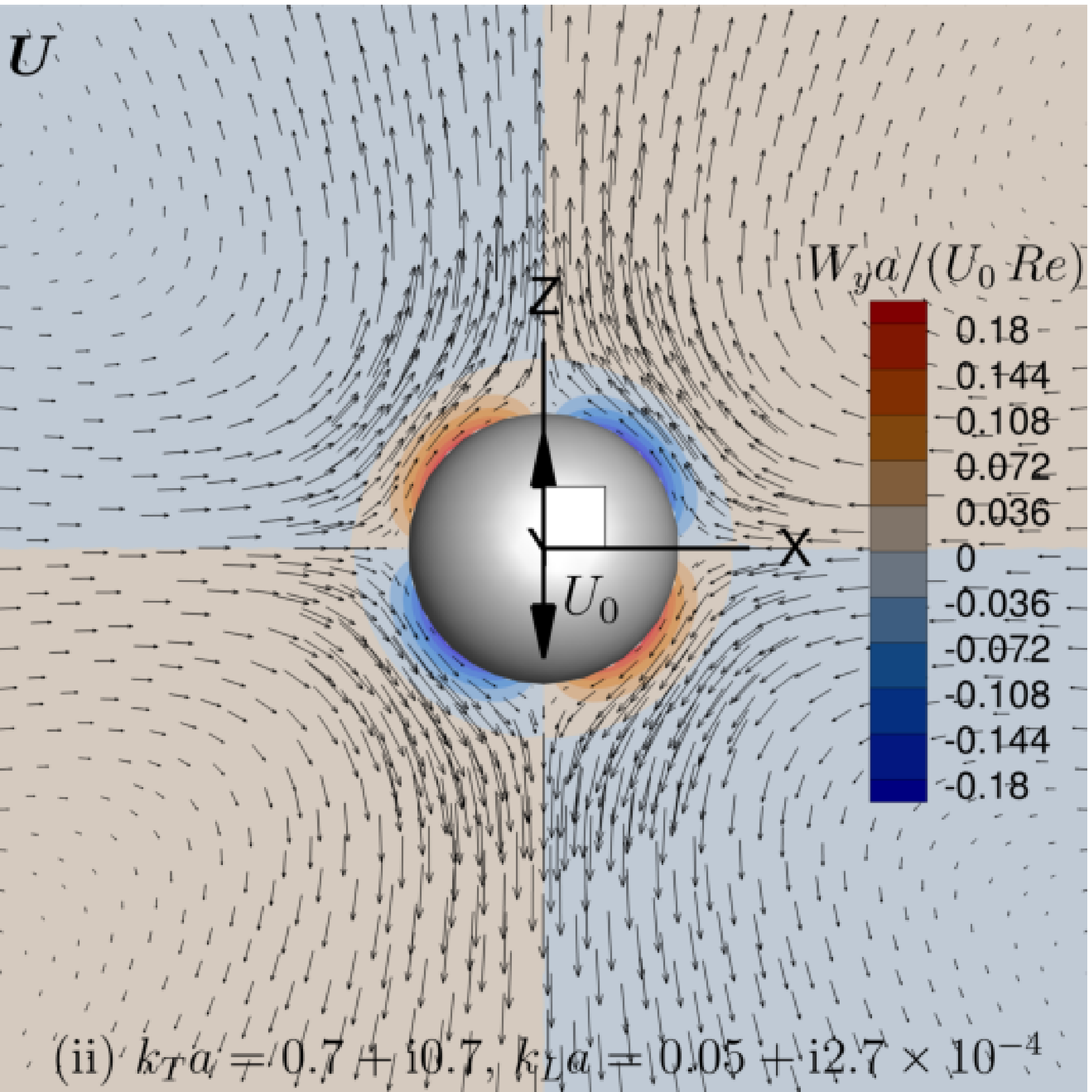}}
\subfloat[]{ \includegraphics[width=0.42\textwidth]{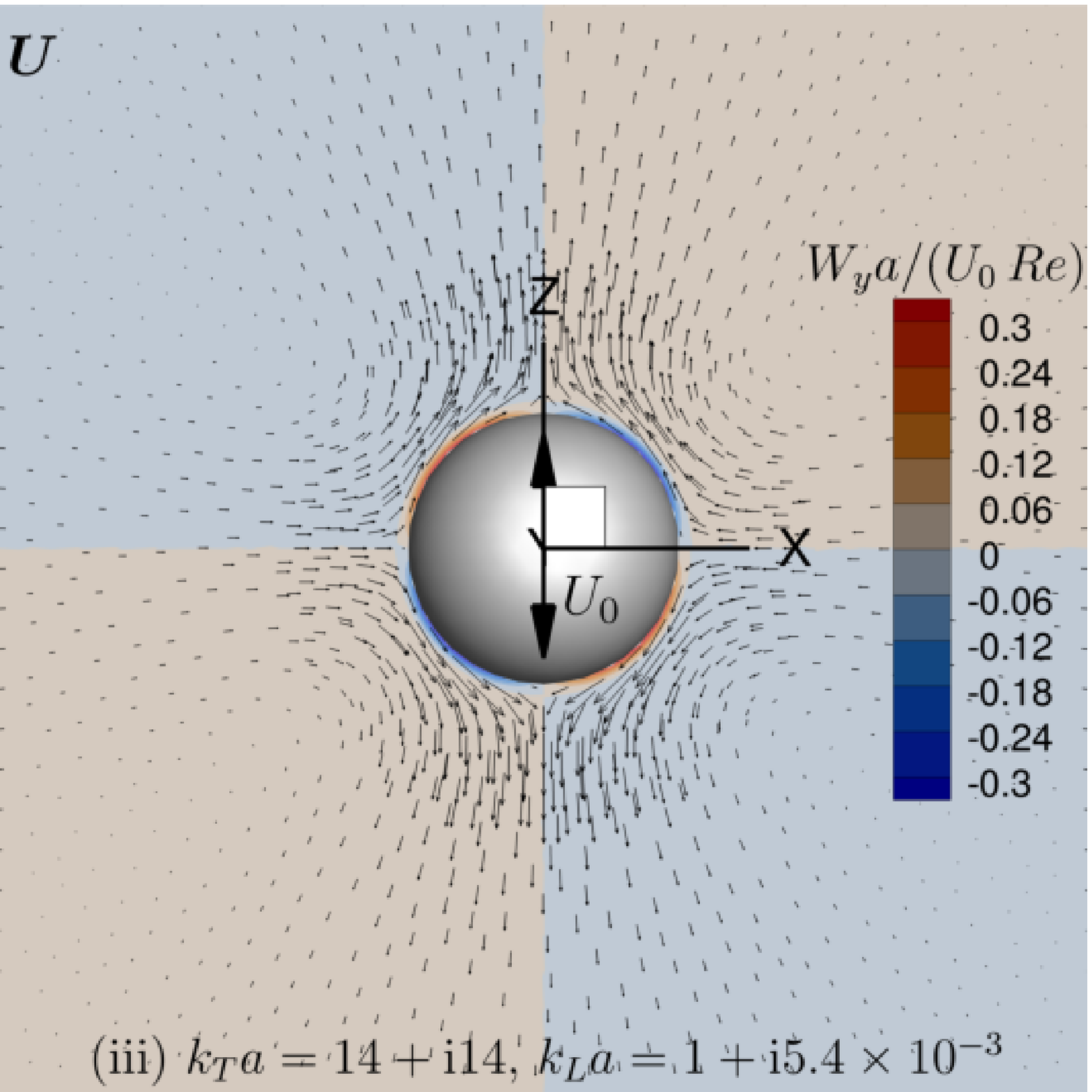}}
\caption{The secondary flow  velocity vector and vorticity distribution on $xz$ plane for (a) Case (ii) with $k_T a = 0.7 + \rmi 0.7$ and $k_L a = 0.05 + \rmi 2.7 \times 10^{-4}$; and (b) Case (iii) when $k_T a = 14 + \rmi 14$ and $k_L a = 1.0  + \rmi 5.4 \times 10^{-3}$.
}  \label{Fig:caseii_iii2}
\end{figure*}

In Fig.~\ref{Fig:vorticity}(a-b), we show results for the function $f(r)$ that characterises the curl of the body force that drives the streaming flow, see (\ref{eq:curl_F}) and the vorticity, $W(r)$ of the streaming flow for the set of $k_Ta$ and $k_La$ values of Cases (i) to (iv) defined in Section~\ref{sec:streaming_examples}. Although the body force is derived from the primary velocity field, it has a much shorter range than the primary velocity field. The streaming vorticity, $W(r)$ has a much shorter range of at most 3 radii from the sphere surface than the streaming velocity components $U_r(r)$ and $U_\theta(r)$ given in Fig.~\ref{Fig:vorticity}(c-d). These velocities are scaled as $U_r/(U_0 Re)$ and $U_\theta/(U_0 Re)$ with the Reynolds number defined as $Re = \rho_0 a U_0/\mu $. Thus the time-averaged body force, $f(r)$ is significant only within a thin boundary layer from the sphere surface and this is reflected in the observation that the vorticity, $W(r)$ is non-zero only within about 5 radii from the surface. However, the streaming velocity field can extend well beyond the range of the streaming vorticity with the typical long ranged characteristics of Stokes flow around a sphere. 

Plots of the secondary velocity, $\boldsymbol U$, and vorticity, $W_y$, are shown in Fig.~\ref{Fig:caseii_iii2} for Cases (ii) and (iii). The four-lobed velocity profile is clearly visible in Fig.~\ref{Fig:caseii_iii2} for the secondary streaming flow. Also clearly visible are the eight regions of the secondary streaming flow vorticity $W_y$ in Fig.~\ref{Fig:caseii_iii2}. Even though the vorticity changes sign in each quadrant, the secondary streaming flow velocity does not reverse direction in an individual quadrant. Near the sphere, the axially symmetric steady acoustic streaming flow, $\boldsymbol U$ due to oscillatory motion of the sphere in the $z$-direction draws fluid towards the sphere in the $xy$-plane and re-directs the fluid away from the sphere symmetrically along the $\pm z$ directions.

\bibliography{AcousticBLref}

\end{document}